\documentclass[11pt]{article}
\usepackage{amssymb}
\usepackage{mathtools}
\usepackage{amsmath}
\usepackage{amstext}
\usepackage{graphicx}
\usepackage{placeins}
\usepackage{epsfig}
\usepackage{verbatim} 
\usepackage{caption}
\usepackage{jcappub}
\usepackage{slashed} 
\usepackage{color}
\usepackage{ulem}

\usepackage{enumitem}
\usepackage{subfigure}
\usepackage{bbm}	
\usepackage{parskip}
\usepackage{dsfont}
\usepackage[numbers,sort&compress]{natbib}

\def\bea{\begin{eqnarray}}
\def\eea{\end{eqnarray}}
\def\be{\begin{equation}}
\def\ee{\end{equation}}
\def\ba{\begin{array}}
\def\ea{\end{array}}

\newcommand{\Mpl}{{M_{\mathrm{Pl}}^2}}

\newcommand{\mpl}{{M_{\mathrm{Pl}}}}

\def\0{{\boldsymbol 0}}

\def\x{{\vec{x}}}

\title{\boldmath The hand-made tail: \\
Non-perturbative tails from multifield inflation}

\author[a,b]{Ana Achúcarro,}
\author[c]{Sebastian Céspedes,}
\author[d]{Anne-Christine Davis,}
\author[e]{Gonzalo A. Palma}

\affiliation[a]{Instituut-Lorentz for Theoretical Physics, Universiteit Leiden,\\ 2333 CA Leiden, The Netherlands}
\affiliation[b]{Department of Physics, University of the Basque Country,\\ 48080 Bilbao, Spain}
\affiliation[c]{Instituto de Física Teórica UAM-CSIC C\\ Nicolás Cabrera 13-15,
Campus de Cantoblanco, 28049 Madrid, Spain}
\affiliation[d]{DAMTP, University of Cambridge, Wilberforce Road, Cambridge, CB3 0WA, UK}
\affiliation[e]{Departamento de Física, FCFM, Universidad de Chile,\\ Blanco Encalada 2008, Santiago, Chile}

\emailAdd{achucar@lorentz.leidenuniv.nl}
\emailAdd{sebastian.cespedes@uam.es}
\emailAdd{ad107@cam.ac.uk}
\emailAdd{gpalmaquilod@ing.uchile.cl}

\abstract{
It is becoming increasingly clear that large but rare fluctuations of the primordial curvature field, controlled by the tail of its probability distribution, could have dramatic effects on the current structure of the universe ---{\it e.g.} via primordial black-holes. However, the use of standard perturbation theory to study the evolution of fluctuations during inflation fails in providing a reliable description of how non-linear interactions induce non-Gaussian tails. Here, we use the stochastic inflation formalism to study the non-perturbative effects from multi-field fluctuations on the statistical properties of the primordial curvature field. Starting from the effective action describing multi-field fluctuations, we compute the joint probability density function and show that enhanced non-Gaussian tails are a generic feature of slow-roll inflation with additional degrees of freedom.}

\numberwithin{equation}{section}
\begin{document}

\maketitle

\section{Introduction}
\label{sec:intro}

The reconstruction of our universe’s history relies on the assumption that the primordial curvature fluctuation $\zeta(\x)$ (responsible for our universe’s inhomogeneities) was initially distributed according to a Gaussian statistics, parametrised by an almost scale invariant power spectrum $P_\zeta(k)$. Although this assumption agrees with every relevant cosmological observation~\cite{Planck:2019kim}, there are good reasons to suspect that our primordial universe could not have been perfectly Gaussian. To start with, the simplest models of cosmic inflation ---the theory that explains the origin of $\zeta(\x)$--- predict tiny, but non-vanishing, levels of non-Gaussianities~\cite{Maldacena:2002vr}. Unfortunately, this minimal prediction will likely remain out of reach for the next generation of cosmological surveys. On the other hand, large non-Gaussianity can arise if, during inflation, $\zeta(\x)$ evolved experiencing large self-interactions and/or interactions with other relevant degrees of freedom~\cite{Cheung:2007st,Chen:2009zp,Chen:2009we,Achucarro:2010da,Achucarro:2012yr}. One way to parametrise the observable effects of these interactions on the distribution of $\zeta(\x)$ is in the form of $n$-point correlation functions $\langle \zeta (\vec x_1) \cdots \zeta (\vec x_n) \rangle$. The shape of these $n$-point functions in momentum space can display distinctive signatures, providing a powerful diagnostic of the types of fields present during inflation. For example, massive fields with spin can leave oscillatory features in the primordial bispectrum (the amplitude of the 3-point correlation function of primordial fluctuations) with a shape determined by their spin~\cite{Arkani-Hamed:2015bza,Lee:2016vti,Achucarro:2018ngj,Arkani-Hamed:2018kmz}.

However, $n$-point correlation functions computed with standard perturbation theory are inappropriate to assess the occurrence of large statistical excursions of $\zeta(\x)$. The prevalence of large statistical excursions is dictated by the shape of the tail of the probability distribution function describing the statistics of $\zeta(\x)$. But perturbative methods fail to correctly determine the profile of tails. As emphasised in ~\cite{Celoria:2021vjw}, perturbation theory schematically relates the 1-point probability distribution of $\zeta$ and connected $n$th-moments $\langle \zeta^n \rangle_c$ [which, in turn, are related to connected $n$-point correlation functions of $\zeta(\x)$] as
\bea
P( \zeta ) &\sim& \exp \left[ - \frac{\zeta^2 }{2 \sigma_\zeta^2}  + \sum_{n=3}^{\infty}  \frac{\langle \zeta^{n} \rangle_c}{\sigma_\zeta^{2 n}} \zeta^{n}   \right]  , \label{P-1-n-moments}
\eea
where $\sigma_\zeta^2$ is the Gaussian variance of the distribution, determined by the power spectrum $P_\zeta(k)$ as $\sigma_\zeta^2 = \int d^3k P_\zeta(k)$. In terms of the usual $f_{\rm NL}$ and $g_{\rm NL}$ parameters for the first few terms in the expansion, the previous expression takes the form
\bea
P( \zeta ) &\sim&  \exp \left[ - \frac{\zeta^2 }{2 \sigma_\zeta^2} \left( 1 +  f_{\rm NL} \zeta  +  g_{\rm NL} \zeta^{2} + \cdots \right) \right] . \label{P-1-few-moments}
\eea
For typical statistical excursions $\zeta \sim \sigma_\zeta \ll 1$, the expansion of the distribution function in terms of moments $\langle \zeta^n \rangle_c$ remains under control as long as $\langle \zeta^n \rangle_c / \sigma_\zeta^n \ll 1$, which can be satisfied in perturbation theory even for values of $f_{\rm NL}$ and $g_{\rm NL}$ of order 1. On the other hand, for unlikely large statistical excursions $\zeta \sim 1$ this expansion may fail, particularly in models predicting $f_{\rm NL}$ and $g_{\rm NL}$ of order 1 (or larger) commonly encountered in theories of inflation involving sizable non-linear interactions (for instance, in the form of interactions with other degrees of freedom). In such models, not only $\langle \zeta^3 \rangle_c$ and $\langle \zeta^4 \rangle_c$, but all $n$-point moments are expected to contribute corrections of order $1$ on the tail of the distribution, making the expansion (\ref{P-1-few-moments}) useless to study extreme fluctuations. This failure of perturbation theory to parametrise large statistical excursions of $\zeta$ in certain models of inflation motivates the consideration of non-perturbative techniques to study the consequence of non-linear interactions of $\zeta$ during inflation~\cite{Flauger:2016idt, Chen:2018brw,Chen:2018uul,Palma:2019lpt,Celoria:2021vjw,Hooshangi:2021ubn,Cai:2021zsp}. 

As unlikely as they might be, large statistical fluctuations of the primordial field $\zeta$ can have dramatic effects on the formation of structure in our universe. More to the point, after inflation, large fluctuations of $\zeta(\x)$ can lead to overdense regions of space that inevitably collapse into primordial black holes (PBHs) (see Refs.~\cite{Carr:2020xqk, Green:2020jor} for recent reviews). These black holes could become the seeds of supermassive black holes at the center of galaxies, and form a substantial part of the dark matter content of our universe. The abundance and clustering properties of these PBHs are extremely sensitive to the shape of the tails of the PDF dictating the distribution of $\zeta(\x)$ ~\cite{Franciolini:2018vbk,Atal:2018neu,Musco:2020jjb,Kitajima:2021fpq}. Thus, to correctly understand the possible generation of PBHs as a result of inflation, we need a reliable, non-perturbative approach to reconstruct the non-Gaussian tails of the primordial fluctuation's PDF. 

The purpose of this work is to quantify precisely the effects of light isocurvature fluctuations on the probability density function of $\zeta(\x)$ by using the non-perturbative approach offered by the stochastic inflation formalism~\cite{Starobinsky:1986fx,Goncharov:1987ir,Salopek:1990re,Starobinsky:1994bd,Tsamis:2005hd,Tolley:2008na,Finelli:2008zg,Finelli:2010sh,Riotto:2011sf,PerreaultLevasseur:2013kfq,Burgess:2014eoa,Moss:2016uix,Grain:2017dqa,Gorbenko:2019rza,Mirbabayi:2019qtx,Cohen:2020php,Pinol:2020cdp}.  
The authors of \cite{Panagopoulos:2019ail} have argued that the interaction between $\zeta$ and a light field $\psi$ can introduce non-Gaussian corrections that modify the shape of tails of the probability density function of $\zeta(\x)$, valid at the end of inflation. Here we confirm this scenario and we show that the joint PDF describing the statistic of $\zeta$ and $\psi$ in two-field models of inflation with canonical kinetic terms is given by
\begin{align}
P(\zeta,\psi) \sim \exp\left[-\frac{\psi^2}{2\sigma_\psi^2}-\frac{1}{2\sigma_\zeta^2}\left(\zeta-\kappa\frac{\psi^2}{2\sigma_\psi^2}\right)^2 + \cdots \right]  ,
\label{two_field_distribution}
\end{align}
where $\kappa$ is related to the strength of the coupling between $\zeta$ and $\psi$, and the ellipses stand for additional subleading contributions that we calculate in some specific examples. The non-perturbative nature of (\ref{two_field_distribution}) is not obvious, but it becomes apparent after integrating over $\psi$ to reveal the tail of the distribution for $\zeta$, which becomes strongly non-Gaussian
\be
P(\zeta)\sim\exp(-\zeta/\kappa) . \label{NG-tail-zeta}
\ee
The dependence on $\kappa$ makes manifest the non-perturbative sensitivity of tails to non-linear interactions between $\zeta$ and other degrees of freedom. Similar non-Gaussian tails have been found in other {\it single-field} scenarios where the background is non trivial~\cite{Ezquiaga:2019ftu,Figueroa:2020jkf,Pattison:2021oen,Figueroa:2021zah}, and quantum diffusion plays an important role. And in~\cite{Panagopoulos:2019ail}, for a sudden, transient coupling between the curvature field and a light spectator field. Instead, in our calculation, slow-roll is preserved throughout and all interactions and couplings are constant in time. In multi-field models of slow-roll inflation with non-canonical kinetic terms we expect corrections in (\ref{two_field_distribution}) that would change the details on how (\ref{NG-tail-zeta}) is obtained, leading to a different profile for the tail. 

An important advantage of the analysis presented here, based on the fluctuations, is that we are able to show that such non-Gaussian tails are a {\it generic} consequence of multifield inflation, and also that they do not require the interruption of slow roll. The leading non-Gaussian contribution to the PDF can be traced back to an ever-present quadratic derivative coupling\footnote{This term is always present unless the inflationary trajectory follows a geodesic in field space} $\dot{\zeta} \psi$ between the curvature and the isocurvature perturbations \cite{Gordon:2000hv, GrootNibbelink:2001qt}. In minimal multifield scenarios the coupling $\kappa$ is related to the angular velocity $\Omega$  of the inflationary trajectory in field space. However,  our results apply to any model where this derivative coupling is present. 

We will be particularly interested in the case of a very light or even {\it massless} (ultralight) isocurvature fluctuation~\cite{Achucarro:2016fby,Achucarro:2019pux}. This provides an alternative inflationary scenario --potentially relevant to  string compactifications-- with predictions currently indistinguishable from those of single-field inflation but where light fields do not need to be stabilized. We consider the background to be quasi de Sitter during the whole duration of inflation. The UV completion of such systems in terms of an effective field space metric and an effective multifield potential has been discussed in \cite{Achucarro:2018ngj,Achucarro:2019pux,Welling:2019bib,Achucarro:2019mea}.

 To derive \eqref{two_field_distribution}, we start from the effective action for the perturbations of a two field model of inflation \citep{Gordon:2000hv,GrootNibbelink:2001qt}. From the action, we will coarse grain the equations of motion to obtain a Fokker-Planck equation for the long wavelength modes. In order to introduce derivative interactions we write the equation in  phase space. Because the time scale associated with the approach to equilibrium of the velocity field $v_\zeta$ is much shorter than the one of the other fields, we can integrate out $v_\zeta$ directly from the Fokker-Planck equation. This leads to Eq.~\eqref{FP_general} which involves only $\psi$ and $\zeta$.  This equation  assumes that the entropy mass of the second field is light, that the  coupling $\Omega^2< H^2$  and that the curvature power spectrum is smaller that one, and to our knowledge has not been previously derived.  Surprisingly enough, it will be possible to solve the time dependent Fokker-Planck equation on a myriad of cases, which among other consequences show that the derivative coupling $\Omega$, both enhances the Gaussian variance and modifies  the PDF introducing a coupling $\kappa\sim\Omega^2/H^2$.

The structure of the paper is as follows. In Section~\ref{sec:SFI} we study the statistics of primordial curvature perturbations and we show how to integrate out $v_\zeta$. We also review some known results in the case of spectator fields on fixed de Sitter. In Section~\ref{sec:linear} we present the linear Fokker-Planck equation for the curvature perturbation coupled to another light field. Since the distribution is Gaussian is possible to obtain exact expressions for the variances of the fields, that as we will show, match known results using standard techniques. Section~\ref{sec:nonlinear} contains the main results of this paper, where we study the full non linear Fokker-Planck equation. After integrating out $v_\zeta$ we will show how it is to obtain \eqref{two_field_distribution} and under which assumptions it holds. Finally, in Section~\ref{sec:conclusions} we conclude and present different ideas to explore in the future. There are a series of appendices where we present technical details of the calculations.

\section{Statistics of primordial curvature perturbations}
\label{sec:SFI}

Before studying the effects of isocurvature fields on the statistics of $\zeta$, we first review the use of the stochastic formalism, showing how it allows a derivation of the probability density function describing the statistics of single fields in quasi-de Sitter backgrounds.  

\subsection{Primordial curvature perturbation}

We start by considering the task of deriving the probability distribution of the primordial curvature perturbation $\zeta$. First, let us recall that the canonical quadratic action for $\zeta$ describing its dynamics during inflation is given by
\begin{equation}
S= \Mpl \int d^4 x a^3\epsilon\left[\dot\zeta^2-\frac{1}{a^2}(\nabla\zeta)^2\right] ,
\end{equation}
where $M_{\rm Pl}$ is the reduced Planck mass. In the previous expression, $a = a(t)$ is the usual scale factor, and $\epsilon$ is the first slow-roll parameter, determined by the Hubble parameter $H = \dot a/ a$ as $\epsilon= - \dot H/ H^2$, and required to be much smaller than 1 throughout inflation. For simplicity, we will disregard slow-roll corrections and take both $H$ and $\epsilon$ as constants. Then, the equation of motion for the long wavelength modes (with wavelengths much larger than the Hubble radius $H^{-1}$) is given by 
\be
\ddot \zeta+3H\dot\zeta = 0.
\ee
One can modify this equation to quantify the influence of short wavelength fluctuations on the evolution of $\zeta$ by introducing a source term representing noise~\cite{Woodard:2005cw}. The resulting equation takes the form
\begin{equation}
\ddot \zeta+3H\dot\zeta=3 H\eta_\zeta ,
\label{eq:pre-LangevinSFI}
\end{equation}
where $\eta_\zeta = \eta_\zeta (t)$ is a time-dependent Gaussian noise with a two-point correlator given by:
\begin{equation}
\langle\eta_\zeta(t)\eta_\zeta(t')\rangle=\frac{H^3}{8 \pi^2\epsilon\mpl^2}\delta(t-t') . \label{correl-eta}
\end{equation}
Equation (\ref{eq:pre-LangevinSFI}) allows one to obtain a Fokker-Planck equation satisfied by the probability density function (PDF) $P(\zeta)$ describing the statistics of long wavelength modes. In order to see this, it is useful to rewrite (\ref{eq:pre-LangevinSFI}) in terms of the following two first order differential equations
\be
\dot\phi_i = \sum_j A_{ij}(t) \phi_j + f_i(t) , \label{eq:LangevinSFI}
\ee
where we have identified $\phi_1 = \zeta$, and $\phi_2 =  \dot \zeta $. Equation (\ref{eq:LangevinSFI}) is a Langevin equation with a drift matrix $A_{ij}$ and noise vector $f_i$ given by
\be
A_{ij} = \begin{pmatrix}
0 & 1 \\
0 & -3H
\end{pmatrix} , \qquad
f_i =  3 H \eta_\zeta  \begin{pmatrix}
0  \\
1
\end{pmatrix} . \label{A-matrix}
\ee
From (\ref{correl-eta}) it follows that the noise vector $f_i$ must satisfy $\langle f_i(t)f_j(t')\rangle= D_{ij}\delta(t-t')$, where $D_{ij}$ is the diffusion matrix, given by
\be
D_{ij} =\begin{pmatrix}
0 &  0\\
0 & D_\zeta
\end{pmatrix} , \qquad D_\zeta \equiv  \frac{9H^5}{8\epsilon \pi^2} . \label{D-matrix}
\ee
In general, one might be interested in computing correlation functions of the stochastic fields fields $\phi_i$ of Eq.~(\ref{eq:LangevinSFI}). These can be computed with the help of a probability density function $P(\phi^i, t)$ derived from the associated Fokker-Planck equation~\cite{van1992stochastic}. The Fokker-Planck equation is determined by $A_{ij}$ and $D_{ij}$ as 
\begin{align}
      \frac{\partial P}{\partial t} + A_{ij} \frac{\partial}{\partial\phi_i}(\phi_j P) - \frac{1}{2}D_{ij} \frac{\partial^2 }{\partial\phi_i\partial\phi_j} P = 0.
\end{align}
Using (\ref{A-matrix}) and (\ref{D-matrix}), we re-express the  Fokker-Planck equation in terms of $\phi_1 = \zeta$ and $\phi_2 = v_\zeta \equiv \dot \zeta$ as
\begin{equation}
\frac{\partial P}{\partial t} + \frac{\partial}{\partial\zeta}\left(v_\zeta P\right) - 3H\frac{\partial}{\partial v_\zeta}(v_\zeta P) - \frac{D_\zeta}{2}\frac{\partial^2}{\partial v_\zeta^2}P = 0. \label{eq:FPcurvature}
\end{equation}
To solve this equation, let us assume a general Gaussian profile of the form
\bea
P(\zeta,v_\zeta,t) &=& 
\frac{1}{2\pi \sqrt{ \det S^{-1} }} \exp\left(- \frac{1}{2} \sum_{ij}  S^{-1}_{ij} \phi_i \phi_j \right) , \label{ansatz-P-coord}
\eea
where $S^{-1}_{ij}$ are the elements of the (symmetric) covariance matrix, whose inverse $S_{ij}$ is constituted by two-point moments as
\be
S_{ij} (t) = 
\begin{pmatrix}
\langle \zeta^2 \rangle (t)  & \langle \zeta v \rangle (t)\\
\langle \zeta v \rangle (t) & \langle v^2 \rangle (t)
\end{pmatrix} .
\ee
The time dependence of $S$ is determined by (\ref{eq:FPcurvature}) together with initial conditions. To determine $S$ we can take the fields $\zeta$ and $v_\zeta$ to be coordinates with Fourier transforms $p$ and $q$ respectively. Then, the Fourier transformed version of \eqref{eq:FPcurvature} is
\begin{align}
\frac{\partial \tilde P}{\partial t} - p \frac{\partial}{\partial q} \tilde P + 3Hq \frac{\partial}{\partial q} \tilde P + \frac{D_\zeta}{2}q^2 \tilde P  = 0 ,
\label{eq:FPFourier}
\end{align}
where $\tilde P$ represents the Fourier transform of $P$. The ansatz given in Eq.~(\ref{ansatz-P-coord}) then implies the following form for $\tilde P$:
\begin{equation}
\tilde P (p, q, t) = \exp\left(- \frac{1}{2} \left[ S_{\zeta \zeta} p^2 + 2 S_{\zeta v} p q +  S_{v v} q^2 \right] \right) . \label{prob-Fourier}
\end{equation}
Replacing this expression back into \eqref{eq:FPFourier} we get the following set of equations satisfied by the elements of the matrix $S$:
\begin{align}
\frac{1}{2}\dot S_{\zeta \zeta} &= S_{\zeta v},  \label{eqs:A-M-3} \\
\dot S_{\zeta v} + 3H S_{\zeta v} &= S_{v v},  \label{eqs:A-M-4} \\ 
\dot S_{v v} + 6H S_{v v} &=  D_\zeta .  \label{eqs:A-M-5}
\end{align}
To solve these equations, we need to impose initial conditions at a given time $t_0$. For instance, consider an initial Gaussian distribution (\ref{ansatz-P-coord}) such that at $t = t_0$ the matrix $S$ contains initial values 
\be
S_{ij} (t_0) = \begin{pmatrix}
S^{(0)}_{\zeta \zeta}  &  S^{(0)}_{\zeta v} \\
S^{(0)}_{v \zeta} & S^{(0)}_{vv}
\end{pmatrix} .
\ee
Solving Eqs.~(\ref{eqs:A-M-3})-(\ref{eqs:A-M-5}) with these initial conditions, we then arrive at: 
\bea
S_{\zeta \zeta}(t) &=& - \frac{2}{3 H} \left(S_{\zeta v}^{(0)} - \frac{D_\zeta}{9 H^2} \right) ( e^{- 3 H (t-t_0)} - 1)+ \frac{1}{9 H^2} \left( S_{v v}^{(0)} - \frac{D_\zeta}{6 H} \right) ( e^{- 6 H (t-t_0)} - 1)
 \nonumber
\\
&& + \left( \frac{2}{3H} S_{v v}^{(0)}  + \frac{D_\zeta}{9 H^2} \right) (t-t_0) + S_{\zeta \zeta}^{(0)} ,
\\
S_{\zeta v}(t) &=& \left(S_{\zeta v}^{(0)} - \frac{D_\zeta}{9 H^2} \right) e^{- 3 H (t-t_0)} - \frac{1}{3 H} \left( S_{v v}^{(0)} - \frac{D_\zeta}{6 H} \right) e^{- 6 H (t-t_0)} ,
\nonumber 
\\
    && + \frac{1}{3H} S_{v v}^{(0)}  + \frac{D_\zeta}{18 H^2} , 
    \\
    S_{v v}(t)&=& \left( S_{v v}^{(0)} - \frac{D_\zeta}{6 H} \right) e^{- 6 H (t-t_0)}  + \frac{D_\zeta}{6 H} .
\eea
The initial values $S_{\zeta \zeta}^{(0)}$, $S_{\zeta v}^{(0)}$ and $S_{vv}^{(0)}$ are the variances associated with long wavelength fluctuations that have already crossed the horizon prior to $t_0$. If we are interested only in the statistics of those modes that cross the horizon starting at $t_0$, we can set the initial values of $S_{ij}$ to $0$. This corresponds to a distribution where the position $\zeta$ and rapidity $v_\zeta$ of the fluctuation are exactly localized at the origin of the field phase space $P(\zeta, v_\zeta, t_0) = \delta(\zeta)\delta(v_\zeta)$. Then, the solutions take the form:
\begin{align}
S_{\zeta \zeta} (t) &=\frac{D_\zeta}{54 H^3}\left(-3+6H(t-t_0)+4e^{-3H(t-t_0)}-e^{-6H(t-t_0)}\right) , \\
S_{\zeta v} (t) &= \frac{D_\zeta}{18H^2}\left(1-e^{-3H(t-t_0)}\right)^2, \\
S_{v v} (z)&= \frac{D_\zeta}{6 H} (1 - e^{- 6 H (t-t_0)}) .
\end{align}
Replacing these expressions back into (\ref{ansatz-P-coord}) we obtain the desired expression for the distribution $P$. The coefficients $S_{\zeta \zeta}$, $S_{\zeta v}$ and $S_{v v}$ depend on time with a characteristic timescale determined by $H^{-1}$. In the limit $t - t_0 \gg H^{-1}$ the distribution simplifies to an asymptotic expression given by
\begin{equation}
P (\zeta, v , t) =\frac{1}{2\pi}\left(\frac{54 H^3}{D_\zeta^2 (t-t_0)}\right)^{1/2}\exp\left[-\frac{9H^2}{2 D_\zeta (t-t_0)}\zeta^2+\frac{3H}{D_\zeta (t-t_0)}\zeta v_\zeta-\frac{6H}{2D_\zeta}v_\zeta^2\right] . \label{eq:SFI_eq_distribution}
\end{equation}
Notice that the widths associated with $\zeta$ and $v_\zeta$ differ in their time dependence, with $v_\zeta$ sharply localized around $0$ (signaling that $v_\zeta$ decays quickly after it becomes super-horizon). In fact, we may marginalize $v_\zeta$ by integrating it from the distribution \eqref{eq:SFI_eq_distribution}, in which case we obtain
\begin{equation}
P(\zeta,t)=\frac{1}{\sqrt{2\pi } \sigma_\zeta }\exp\left(-\frac{1}{2 \sigma_\zeta^2}\zeta^2\right)  ,
\label{PDF:zeta}
\end{equation}
which is a Gaussian distribution for $\zeta$ with variance $\sigma_\zeta^2$ given by
\be
\sigma_\zeta^2 = \frac{D_\zeta}{9 H^2} (t - t_0) =  \frac{H^3 }{8\epsilon \pi^2} (t - t_0) .
\ee
Recall that this expression is valid for $t \gg H^{-1}$, provided the initial condition $S_{ij} = 0$ at $t_0 = 0$. The time dependence of the variance $\sigma_\zeta^2$ just reflects the fact that as time progresses, more and more modes populate the long wavelength regime. In this way, at a given time $t$, the probability distribution $P(\zeta,t)$ describes the statistics of long-wavelengths that crossed the horizon between $t_0$ and $t$.

\subsection{Integrating out $v_\zeta$} 
\label{sec:integrat-v_zeta}

Although $\zeta$ and $v_\zeta$ had the same status in the treatment leading to \eqref{eq:SFI_eq_distribution} we are ultimately interested only in the statistics of $\zeta$ (after all, $v_\zeta$ decays quickly on super-horizon scales). This led us to derive \eqref{PDF:zeta} after marginalizing $v_\zeta$. Alternatively, we can integrate out $v_\zeta$ at an early stage, and obtain a Fokker-Planck equation only for $\zeta$. For instance, if we neglect the second derivative of $\zeta$ in \eqref{eq:LangevinSFI}, the system reduces to a single Langevin equation $\dot\zeta=\eta_\zeta$, yielding the following Fokker-Planck equation
\begin{align}
    \frac{\partial P}{\partial t}=\frac{D_\zeta}{18H^2}\frac{\partial^2 P}{\partial \zeta^2} ,
\label{zeta_FP}
\end{align}
whose solution is precisely given by Eq.~\eqref{PDF:zeta}. Notice that \eqref{eq:FPcurvature} included a term involving a second derivative with respect to $v_\zeta$, and only a first derivative with respect to $\zeta$. In contrast, Eq.~\eqref{zeta_FP} contains a second derivative with respect to $\zeta$. This can be understood as the effect of integrating out $v_\zeta$ over the Fokker-Planck equation~\eqref{eq:FPcurvature} with the ideas of Refs.~\cite{VanKampen:1985,van1992stochastic}, which go as follows: First, by noticing that the time scale in which $v_\zeta$ becomes time independent is given by $t_v=1/3H$, we can rewrite Eq.~\eqref{eq:FPcurvature} as:
\begin{align}
  \frac{\partial}{\partial v_\zeta}\left(v_\zeta+\frac{t_v D_\zeta}{2 }\frac{\partial}{\partial v_\zeta}\right)P = t_v\left(\frac{\partial}{\partial t}+v_\zeta\frac{\partial}{\partial \zeta}\right)P.
  \label{rewritten_FP}
\end{align}
Here, the terms on the right hand side (RHS) are much smaller than those on the left hand side (LHS). Indeed if we assume that the fields are given by their typical values $\zeta\sim\sqrt{\sigma_\zeta^2}$ then we have that the term containing the time derivative is of order $t_v\dot\sigma^2_\zeta P\sim\Delta_\zeta^2P$ while the second term is of order $1/\sqrt{Ht}P$. On the other hand the terms on the LHS are both  of order $\mathcal{O}(1) \times P$. We can make use  of this hierarchy if we expand the PDF in powers of $H t_v$:
\begin{align}
    P(\zeta,v_\zeta,t)=P^{(0)}+Ht_v P^{(1)}+(Ht_v)^2P^{(2)}+\dots .
\end{align}
Then, by replacing this expression back into~\eqref{rewritten_FP} we find that at leading order in $t_v$ the equation for the first term $P^{(0)}$ is simply given by 
\begin{align}
    \frac{\partial}{\partial v_\zeta}\left(v_\zeta+\frac{t_v 2D_\zeta}{2 }\frac{\partial}{\partial v_\zeta}\right)P^{(0)} =0 ,
\end{align}
whose solution can be written as
\begin{align}
    P^{(0)}(\zeta,v_\zeta;t) = e^{-\frac{6H}{2D_\zeta}v_\zeta^2}\phi_0(\zeta,t) ,
\end{align}
where $\phi_0$ is a function of $\zeta$ and $t$ that can be determined by considering the equation for $P^{(1)}$:
\begin{align}
     \frac{\partial}{\partial v_\zeta}\left(v_\zeta+\frac{t_v D_\zeta}{2 }\frac{\partial}{\partial v_\zeta}\right)HP^{(1)}  = \left(\frac{\partial \phi_0}{\partial t}+v_\zeta \frac{\partial\phi_0}{\partial\zeta}\right)e^{-\frac{6H}{2D_\zeta}v_\zeta^2} . \label{P1}
\end{align}
Now, notice that in this last equation the LHS corresponds to a total derivative of $v_\zeta$, whereas the RHS is proportional to Gaussian function of $v_\zeta$. Hence we can integrate it with respect to $v_\zeta$ and obtain the constraint equation $\frac{\partial \phi_0}{\partial t}=0$. This in turns allows us to write a solution for $P_1$ given by
\begin{align}
    P^{(1)}(\zeta,v_\zeta;t)=- \frac{v_\zeta}{H}\frac{\partial\phi_0}{\partial\zeta}e^{-\frac{6H}{D_\zeta}v_\zeta^2}+\frac{\phi_1(\zeta,t)}{H}e^{-\frac{6H}{2D_\zeta}v_\zeta^2} ,
\end{align}
which now depends on another function $\phi_1(\zeta,t)$. Repeating the same step we can find, at the next order in $t_v$, the following equation for $P^{(2)}$:
\begin{align}
     \frac{\partial}{\partial v_\zeta}\left(v_\zeta+\frac{t_v D_\zeta}{2 }\frac{\partial}{\partial v_\zeta}\right)H^2P^{(2)}(\zeta,v_\zeta,\psi;t)=\left(\frac{\partial \phi_1}{\partial t}+v_\zeta \frac{\partial\phi_1}{\partial\zeta}+v_\zeta^2\frac{\partial^2\phi_0}{\partial_\zeta^2}\right)e^{-\frac{6H}{2D_\zeta}v_\zeta^2} .
     \label{P2}
\end{align}
Given that this equation has the same structure as~\eqref{P1}, we immediately infer, after integrating over $v_\zeta$, that
\begin{align}
    \frac{\partial\phi_1}{\partial t}=\frac{D_\zeta}{6 H}\frac{\partial^2\phi_0}{\partial_\zeta^2} .
    \label{constraint_tv2}
\end{align}
Collecting the terms for the PDF we obtain,
\begin{align}
    P(\zeta,v_\zeta,t)= \Big[ \phi_0(\zeta) + t_v v_\zeta\frac{\partial\phi_0}{\partial\zeta} +t_v\phi_1(\zeta,t) \Big] e^{-\frac{6H}{2D_\zeta}v_\zeta^2} .
\end{align}
After integrating over $v_\zeta$, this result reduces to 
\begin{align}
    P(\zeta ;t)=\phi_0(\zeta)+t_v\phi_1(\zeta,t) .
\end{align}
Finally, using the constraint equations for $\phi_0$ and $\phi_1$ we find, up to first order in $t_v$, that $P(\zeta ;t)$ must satisfy
\begin{align}
    \frac{\partial }{\partial t}P(\zeta,t)=\frac{D_\zeta}{18H^2}\frac{\partial^2 P(\zeta,t)}{\partial\zeta^2} ,
\end{align}
which is the Fokker-Planck equation~\eqref{zeta_FP} obtained by ignoring the $\ddot\zeta$ term in the Langevin equation. Going beyond second order in $t_v$ does not add any further correction as the equations obtained for the other terms in the PDF expansion are the same as those in \eqref{P2}.

\subsection{Spectator fields in de Sitter}

To complement the previous discussions, we now study the statistics of a light spectator field on a de Sitter background. The light scalar field $\psi$ has potential $V(\psi)$ and its equation of motion is given by
\begin{align}
\ddot\psi+3H \dot \psi+\frac{k^2}{a^2}\psi+V'(\psi)=0 .
\end{align}
We take $V''(\psi)\ll H^2$ since the field is light. As we did with $\zeta$, the statistical properties of $\psi$ can be studied by dividing the field into long and short wavelength modes. In this way, the long wavelength field $\psi$, satisfies the Langevin equation
\begin{align}
\dot \psi=-\frac{V'(\psi)}{3H}+\eta_\psi(t) , \label{Langevin-psi}
\end{align}
where $\eta_\psi$ is a Gaussian noise representing the effects of the short wavelength modes. The correlation function of the noise term is given by
\begin{align}
\langle\eta_\psi(t)\eta_\psi(t')\rangle=\frac{H^3}{4\pi^2}\delta(t-t').
\end{align}
Notice that we have chosen to disregard the role of $\ddot \psi$ in the Langevin equation, which can be justified with the same arguments given in Section~\ref{sec:integrat-v_zeta}. From the Langevin equation (\ref{Langevin-psi}) it is possible to compute the one point probability distribution function $P(\psi,t)$ by writing  the associated Fokker-Planck equation:
\begin{align}
\frac{\partial P}{\partial t}=\frac{1}{3H}\frac{\partial}{\partial\psi}\left(V'(\psi)P\right)+\frac{H^3}{8\pi^2}\frac{\partial^2P}{\partial\psi^2} .
\label{eq:FP_example}
\end{align}
This equation is highly non linear since  the drift $-\frac{V'(\psi)}{3H}$ is an arbitrary function of $\psi$. Nevertheless it is possible to find an exact solution. This is done by first noticing that \eqref{eq:FP_example} has an equilibrium solution
\begin{align}
\lim_{t\to\infty}P(\psi,t)=\exp\left(-\frac{8\pi^2 V(\psi)}{3H^4}\right) ,
\label{equlibrium_distributrion}
\end{align}
 which is obtained by imposing that $P$ is time independent. To obtain solutions to \eqref{eq:FP_example} we can now write $P(\psi,t)$ as
\begin{align}
P(\psi,t)=\exp\left(-\frac{4\pi^2 V(\psi)}{3H^4}\right)\sum_{n=0}^{\infty}a_n\Phi_n(\psi)e^{-\Lambda_n(t-t_0)},
\label{eq:FP_decomposition}
\end{align}
where the coefficients $\Lambda_n$ and the functions $\Phi_n$ satisfy the following eigenvalue problem
\begin{align}
\left(-\frac{1}{2}\frac{\partial^2}{\partial\psi^2}+\frac{1}{2}(v'(\psi)^2-v''(\psi))\right)\Phi_n(\psi)=\frac{4\pi^2\Lambda_n}{H^3}\Phi_n(\psi),\qquad v(\phi)\equiv \frac{4\pi^2}{3H^4}V(\psi).
\label{eq:eigenvalue_FP}
\end{align}
The time dependence of $P(\psi, t)$ is controlled by the eigenvalues $\Lambda_n$, which are positive and, for general potentials $V(\psi)$, their value increase with $n$ (with $\Lambda_0=0$). This implies that the decay rate to the equilibrium distribution is  given by $1/\Lambda_1$. For instance, when the potential is quadratic ($V(\psi)=\frac{1}{2}m^2\psi^2$)  one finds the  solution of \eqref{eq:eigenvalue_FP} is given by Hermite polynomials  with eigenvalues given by  $\Lambda_n= \frac{m^2}{3H^2}\times n$. In this case, the solution reaches equilibrium for $\Delta N \gg H^2/m^2\gg 1$. 

Using the decomposition \eqref{eq:FP_decomposition} it is also possible to deduce the statistical properties of the equal time correlation function $G(R)$, where $R$ is the distance between two points, found as
\be
G(R)=N\sum_n\vert A_n\vert^2e^{-2\log(RH)\Lambda_n/H},
\ee
where the coefficients $A_n$ are given by
\be
A_n=N^{-1}\int d\psi \psi e^{-\frac{4\pi^2}{3H^4}V(\psi)}\Phi_n(\psi) .
\ee
From this result, we may conclude that the stochastic approach is valid for a patch of size 
$R\sim H^{-1}e^{H/\Lambda_1}$.  For a quadratic potential this is of order $R\sim H^{-1}e^{H^2/m^2}\gg H^{-1}$, which implies that the statistical average occurs over a  large number of Hubble patches. This quantity has to be compared with the correlation length of the observed universe, given by $\sim H^{-1}e^{\Delta N}$. This implies that a field  fits inside the observed universe if $\Delta N<H^2/m^2$.

The present analysis assumed a fixed de Sitter background, but it can be generalised to the case of quasi-de Sitter backgrounds, as required to study slow-roll inflation. In this case there are added difficulties. One problem involves the role of gauge transformations on Hamiltonian constraints satisfied at the level of the Langevin equations. In the following discussion we will avoid this issue by assuming that the graviton is decoupled from scalar fields ({\it i.e.} we consider the decoupling limit, in which the mixing with gravity is negligible for energies larger than $\sqrt{\epsilon}H$). Furthermore we will assume that the time dependence of the couplings is negligible over the time scales we will consider (typically an $e$-fold). Within this regime, the dynamics reduces to study the action for the curvature perturbation $\zeta$ coupled to an isocurvature field $\psi$ via derivative couplings. In this way the problem is analogous to studying two coupled spectator fields, evolving on de Sitter.

\section{Statistics for two-field inflation}
\label{sec:linear}

In this section we use the tools introduced in the previous section to derive the probability density function describing the statistics of fluctuations in multifield theories. For now, we shall restrict our treatment to the case of two-field models, and focus on the case of theories with linear interactions. In Section~\ref{sec:nonlinear} we consider the role of non-linear interactions. 

Our starting point is to consider the two-field action (background plus perturbations) describing inflation:
\be
S = S_{\rm EH}  - \frac{1}{2} \int d^4 x \sqrt{- g} \left[ \gamma_{ab} (\phi) \partial \phi^a \partial \phi^b + V(\phi)  \right] .
\ee
where $S_{\rm EH}$ is the Einstein-Hilbert action, $\gamma_{ab} (\phi)$ is a sigma model metric describing the geometry of the scalar field target space and $V(\phi)$ is the scalar potential driving inflation. Deviations from a geodesic trajectory are parametrised by the angular velocity $\Omega$. The action for the curvature field $\zeta$ and the isocurvature field $\psi$ is obtained by decomposing 
the fields
along tangent and normal directions to the inflationary trajectory  (see \cite{Gordon:2000hv, GrootNibbelink:2001qt,Achucarro:2010jv,Achucarro:2010da} for a more detailed explanation).
 The quadratic action is given by~\cite{Achucarro:2016fby}
\begin{equation}
S=\frac{1}{2}\int d^4 x a^3\left[f_\zeta^2\left(\dot\zeta-\frac{2\Omega}{f_\zeta}\psi\right)^2-\frac{f_\zeta}{a^2}(\nabla\zeta)^2+\dot\psi^2+\frac{1}{a^2}(\nabla\psi)^2+\mu^2\psi^2\right] , \label{Action-two-field-linear}
\end{equation}
where $\Omega$ is the coupling between the two fields, and $\mu$ is the so called entropy mass of $\psi$. In addition, we use $f_\zeta^2\equiv 2\mpl^2|\dot H|/H^2=2\epsilon\mpl^2$. The equations of motion resulting from the variation of the previous action are
\begin{align}
\ddot \zeta+3H\dot\zeta+\frac{k^2}{a^2}\zeta&=-\frac{2\Omega}{f_\zeta}(\dot\psi+3H\psi) ,  \\
\ddot\psi+3H\dot\psi+\frac{k^2}{a^2}\psi+m^2\psi&=2\Omega f_\zeta\dot\zeta ,
\label{eq:MF}
\end{align}
where\footnote{As discussed in Refs.~\cite{Achucarro:2012yr,Achucarro:2018ngj}, $\mu$ is the physical mass that identifies the rest energy of one of the quanta in the spectrum of the theory on subhorizon scales, whereas $m$ is just a mass parameter entering the equation of motion \eqref{eq:MF}. The quantity $m^2$ can be large and negative without affecting the stability of the system, as long as $\mu^2\geq 0$.} $m^2=\mu^2-4\Omega^2$. Notice that the coupling $\Omega$ mixes both fields, but these equations can still be solved using perturbation theory if we assume that the coupling satisfies  $\Omega/H\ll 1$. In this case we have that the two point functions for $\zeta$ and $\psi$ at horizon crossing, are given by
\begin{align}
\Delta_\zeta^2=\frac{1}{8\pi^2\mpl^2}\frac{H^2}{\epsilon},\qquad\Delta_\psi^2=\frac{H^2}{4\pi^2} .
\label{scale_invariant_ps}
\end{align}
Since  $\psi$ is a light field, it can continue evolving after crossing the horizon. When the coupling $\Omega \neq 0$, $\psi$ seeds the perturbations of $\zeta$ which could lead to its correlation functions growing on superhorizon scales~\cite{Achucarro:2016fby}. As we will see all these effects can be properly incorporated by studying the linearised Langevin equations. 

In order to analyse the stochastic dynamics let us note that $\psi$ couples to $v_\zeta$. In order to include these terms it is more convenient to use the phase space formulation of the Langevin equations. As explained in Section~\ref{sec:SFI} this is achieved by introducing the time derivatives of the fields in the Langevin equations. For \eqref{eq:MF} these correspond to
\begin{align}
\dot\zeta&=v_\zeta,\nonumber\\
\dot v_\zeta&=-3Hv_\zeta-2\frac{\Omega}{f_\zeta}v_\psi-\frac{6\Omega H}{f_\zeta} \psi+\eta_\zeta,\nonumber\\
\dot \psi&=v_\psi,\nonumber\\
\dot v_\psi&=-3Hv_\psi-{m^2}\psi+2\Omega f_\zeta v_\zeta+\eta_\psi ,
\label{eq:linearisedLangevin}
\end{align}
where $\eta_\zeta$ and $\eta_\psi$ are Gaussian noise terms with correlation functions given by
\begin{align}
\langle\eta_\zeta(t)\eta_{\zeta}(t')\rangle=H^2\Delta_\zeta^2\delta(t-t'),\qquad\langle\eta_\psi(t)\eta_{\psi}(t')\rangle=\frac{H^3}{4\pi^2}\delta(t-t') .
\end{align}
Notice that the  Langevin  equations are coupled, hence the  PDF do not factorise into $P\propto P_\zeta(\zeta,v_\zeta)P(\psi,v_\psi)$. This adds some complications due to the fact that $\zeta$ does not reach equilibrium, and so it is not possible to {\it a priori} make use of the decomposition \eqref{eq:FP_decomposition} to find the PDF. 

Nevertheless, since the Langevin equations are linear it is possible to find an exact  Gaussian solution. In this case all we need to do is to compute the covariance matrix, as explained in Appendix~\ref{sec:General_solution}. Since  we have the drift and the noise matrices given by
\begin{align}
A=\left(
\begin{array}{cccc}
 0 & 0 & 1 & 0 \\
 0 & 0 & 0 & 1 \\
 0 & -\frac{6H\Omega}{f_\zeta} & -3 H & -\frac{2\Omega}{f_\zeta} \\
 0 & -m^2 & 2f_\zeta \Omega & -3 H \\
\end{array}\right)\label{A_mf_matrix},\qquad D=\left(
\begin{array}{cccc}
 0 & 0 & 0 & 0 \\
 0 & 0 & 0 & 0 \\
 0 & 0 & D_\zeta & 0 \\
 0 & 0 & 0 & \frac{9 H^5}{4 \pi ^2} \\
\end{array}
\right),
\end{align}
the covariance matrix is given by
\begin{align}
C(t)=\int_0^t\exp({(t-t')A})D \exp((t-t')A^{t}) dt' \ ,\label{eq:cov_matrix}
\end{align}
where we have assumed initial conditions given by $P=\delta(\zeta)\delta(v_\zeta)\delta(\psi)\delta(v_\psi)$ and we have set the initial time to zero. Notice that it is not necessary to write the explicit PDF since the variances are given by
\begin{align}
\sigma^2_{\phi_a}=C_{\phi_a \phi_a},
\end{align}
where $\phi_a$ is one of the fields and also the corresponding element on the diagonal of $C$. Off diagonal elements of $C$ are cross correlations between different fields.
Before writing explicit expressions for $C(t)$ let us notice that the coupled dynamics imply that there are several time scales over which the field decays. A useful way of understanding this is by noticing that the  time dependence is encoded in the exponential of the drift matrix $A$. Since for $\mu\neq 0$, $A$ is diagonalisable it can be written as $A=U D U^{-1}$ with $D$ a diagonal matrix containing the eigenvalues $\lambda_i$ of $A$. Using this decomposition, the exponential of the drift matrix can be written as $\exp(At)=U\exp(D t)U^{-1} $ which implies that the time dependence will appear in terms containing $e^{\lambda_i t}$.

\begin{figure}
  \includegraphics[scale=0.45,trim=1.8cm 12cm 1cm 4cm,clip]{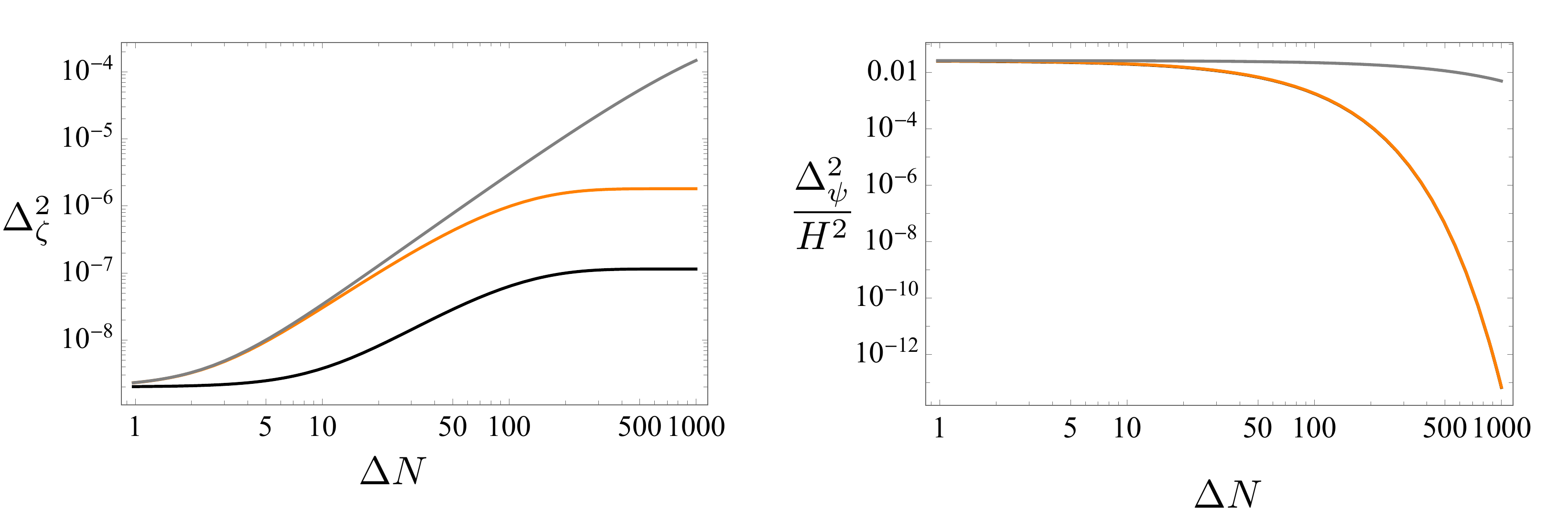}
\caption{We plot the scale invariant power spectra  \eqref{time_dependent_ps-1} amd \eqref{time_dependent_ps-2}. We set $\Delta_\zeta^0=2\times 10^{-9}$. The parameters for the black line are $\mu=0.2 H$, $\Omega=.05$, the orange line $\mu=0.2 H$, $\Omega=.2H$ and for the gray line $\mu=.05H$, $\Omega=.2H$}
\label{Fig:ps}
\end{figure}

For the drift matrix $A$ given in \eqref{A_mf_matrix} the eigenvalues $\lambda_i$  are, $0$,$-3H$, $-3H(1+\mu^2/H^2)$ and $-\mu^2/3H$. Since  the integrand will contain factors of $\exp t(\lambda_i+\lambda_j) $  there are three main cases. First if $\lambda_i=\lambda_j=0$  after integrating implies the appearance of terms linear  in $t$, which are due to  the variance of $\zeta$ always growing with time. The others  cases arise when one of the eigenvalues  is different from zero. When the sum of the eigenvalues is proportional to $3H(1+\mu^2/H^2)$ the term decays after a time $t\sim 1/(3H)$ analogous to $v_\zeta$ in the single field case. After integrating this term will generate two pieces, one is constant because it is evaluated at $t'=t$ and the other one decays. The third case is when the sum of the eigenvalues is proportional to  $\mu^2/H$. These terms decay on a longer time scale dictated by the isocurvature mass  $t_\psi\sim H/\mu^2$. Notice, moreover, that the decay depends  on the isocurvature mass, not on $m$. As an example  let us look at the variance of $\zeta$. From \eqref{eq:cov_matrix}, this quantity is found to be given by
\begin{align}
\sigma_\zeta^2=C_{\zeta\zeta}\approx H t \Delta_\zeta^2\left(1+\frac{36 H^2\Omega^2}{\mu^4}\right) -54 \Delta_\zeta^2\frac{H^4\Omega^2}{\mu^6}\left(3-4e^{-\frac{\mu^2}{3H}t}+e^{-\frac{2\mu^2}{3H}t}\right),
\label{eq:variance_zeta}
\end{align}
where we have kept only leading order terms and we have assumed  $Ht\gg 1$. As anticipated, the time dependence appears both linearly in $t$ and  in powers of $\exp(-\mu^2 t/H)$. The first dependence is due to the modes entering the comoving horizon as we discussed in Section~\ref{sec:intro}. Notice, however, that  there is a new piece which depends on the mass and the coupling of the extra field. The second term is due to the superhorizon evolution of the light scalar field which takes a longer time to decay. At very large times $t\gg t_\psi$ this contribution is suppressed and the time dependence is as usual.  However, at intermediate times  $1\ll H t\leq H^2/\mu^2$ the time dependence from the exponential can dominate. To see this, let us expand the variance in small $\mu$
\begin{align}
\sigma_\zeta^2=C_{\zeta\zeta}\approx \Delta_\zeta^2 Ht\left(1+\frac{4}{3}(Ht)^2\frac{\Omega^2}{H^2}\right)+\Delta_\zeta^2\Omega^2/H^2\times\mathcal O( (Ht)^4\mu^2/H^2).
\end{align}
There are  now  cubic terms  in $Ht$, which do not depend on the entropy mass but only on the factor $\Omega/H$. This effect is due to the light field behaving as a massless ({\it ultralight}) field before it settles into an equilibrium distribution. This can be seen more  directly if we look at the variance of $\psi$ for $H t \gg 1$
\begin{align}
\sigma_\psi^2=C_{\sigma\sigma}\approx\frac{3H^4}{8\pi^2\mu^2}\left(1-e^{-\frac{2\mu^2}{3H}t}\right) .  \label{eq:variance_psi}
\end{align}  
At late times the exponential can be neglected and one recovers the equilibrium value. Interestingly at intermediate times  $1\ll H t\leq H^2/\mu^2$ , expanding in small $\mu$ we find
\begin{align}
\sigma_\psi^2\approx  \frac{H^3 t}{4\pi^2}-\frac{H^2 t^2}{12\pi^2}\mu^2, 
\end{align}
which implies that for small $(H t) \mu^2/H^2$ the field behaves as a massless field. Later, when the second piece becomes of order $(H t)\mu^2/H^2\sim 1$ or larger this expansion stops being valid and we need to consider the full expression. It is also interesting to rewrite  the variances in terms of the scale invariant power spectrum. This can be done by writing time in terms of the number of efolds after horizon crossing.  If a mode with wavenumber $k$ leaves the horizon at a time $t_k$ we can then write $H(t_k+t_*)=\log(k/k_*)$ where $k_*$ and $t_*$ are the longest scale and time measured. Since the variance of a field $\phi$ is computed as
\begin{align}
\sigma_\phi^2=\int \frac{dk}{k} \Delta_\phi^2 ,
\end{align} 
it is possible  to invert this relation and write $\Delta_\phi$ as the  logarithmic derivative of the variance with respect to $k$. It is also convenient to express all quantities in terms of e-folds using $\Delta N=H(t_k+t_*)$. Finally we find that
\begin{align} 
\Delta_\zeta^2(t)&=\frac{H^2}{8\pi^2\mpl^2\epsilon}\left(1+36\frac{H^2\Omega^2}{\mu^4}\left(1-e^{-\frac{\mu^2}{3H^2}\Delta N}\right)^2\right) , \label{time_dependent_ps-1} \\
\Delta_\psi^2(t)&=\frac{H^2}{4\pi^2}e^{-\frac{2\mu^2}{3H^2}\Delta N} .
\label{time_dependent_ps-2}
\end{align}
Notice that at $t=0$ we have $\Delta_\zeta^2 (0) = \Delta_\zeta^2$, where $\Delta_\zeta^2$ is defined in (\ref{scale_invariant_ps}). That is, the value of $\Delta_\zeta^2$  at horizon crossing acts as an initial condition for $\Delta_\zeta^2 (t)$, which continues to evolve since the light field has not yet reached equilibrium. Notice that similar formulae were obtained by using the in-in formalism in~\cite{Achucarro:2019mea}. From the last formulae we also see that, in general, light fields behave as massless fields during a time $1\ll \Delta N\ll 3H^2/\mu^2$. Expanding again in small $\mu$ we find that
\begin{align}
\Delta_\zeta^2 (t)&=\frac{H^2}{8\pi^2\mpl^2\epsilon}\left(1+4(\Delta N)^2\frac{\Omega^2}{H^2}\right) .
\end{align}
Notice that this is same superhorizon  growth described in the case of ultralight fields \cite{Achucarro:2016fby}; this is the case when the entropy mass is exactly zero. We plot \eqref{time_dependent_ps-1} and \eqref{time_dependent_ps-2} in Fig.\ref{Fig:ps}. We  see  that   for curvature field there is an initial ultralight phase which will last as long as $\mu^2\Delta N/3H^2\ll 1$. Notice that if this inequality holds until the end of inflation then $H^2/\mu^2\ll \Delta N$, the ultralight phase is all there is, since the field does not have time to start moving away from a massless distribution. Another feature which we have previously discussed is that the coupling changes the final amplitude of the primordial fluctuation. In the case of the light field we see that the power spectrum eventually decays to zero

\subsection{Probability distribution function}
\label{sec:pdf}

Having understood the time dependence of the variances, now we would like to derive the PDF of the two-field model. To start with, let us notice that the time scales associated with the velocity fields $t_{v_\psi}$ and $t_{v_\zeta}$ are much smaller than the scale of $\zeta$ and $\psi$. Using this it is possible to integrate out $v_\psi$ and $v_\zeta$ from the Fokker-Planck equation. We will explain in detail how this is done in Section~\ref{Adiabatic_elimination}. The resulting Fokker-Planck equation is
\begin{align}
\frac{\partial P}{\partial t}=-\frac{2\Omega}{f_\zeta}\frac{\partial}{\partial\zeta}(\psi P)+\frac{2\Delta_\zeta^2\Omega}{H}\frac{\partial^2}{\partial\zeta\partial\psi}(\psi P)+\frac{H\Delta_\zeta^2}{2}\frac{\partial^2P}{\partial\zeta^2}+\frac{\partial}{\partial\psi}\left(t_\psi^{-1}\psi P+\frac{D_\psi}{2}\frac{\partial P}{\partial\psi}\right) ,
\end{align}
where 
\begin{align}
D\psi\equiv\frac{H^3}{4\pi^2},\qquad t_\psi\equiv \frac{3H}{\mu^2} .
\end{align}
This Fokker-Planck equation is still linear, even though it has a mixed noise term, and it can be solved with the techniques of Appendix~\ref{sec:General_solution}. Further simplification is possible if we consider the following: As explained in Appendix~\ref{sec:General_solution}, the noise term is computed by  using the two-point function of the field at horizon crossing. Nevertheless, we have seen that the value of the variance grows with time on superhorizon scales. On the other hand, a direct computation of the two point function also shows a superhorizon growth~\cite{Achucarro:2016fby,Achucarro:2019pux}
\begin{align}
\Delta_\zeta^2(t)=\frac{H^2}{8\pi^2\mpl^2\epsilon}\left(1+36\frac{H^2\Omega^2}{\mu^4}(1-e^{-t/t_\psi})^2\right) ,
\end{align}
which, of course, is the same result we obtained in \eqref{time_dependent_ps-1}. This can be understood as follows. As it has been previously pointed out in Refs.~\cite{Achucarro:2016fby,Achucarro:2019pux}, in the long wavelength limit the equations of motion for $\zeta$ decouple if they are written in terms of $v_\zeta=\dot\zeta-\frac{2\Omega}{f_\zeta}\psi$
\begin{align}
\dot v_\zeta+3Hv_\zeta+\frac{k^2}{a^2}=0 , \\
\ddot\psi+3H\dot\psi+\frac{k^2}{a^2}\psi+\mu^2\psi=2\Omega f_\zeta v_\zeta .
\end{align}
Notice that in the long wavelength limit the first equation admits as a solution $v_\zeta=v_\zeta^0 a^{-3}$. Plugging this solution back into the second equation implies that the last source term vanishes for $t\gg 3H$, hence both equations are decoupled. Nonetheless, the fact that $v_\zeta$ also depends on $\psi$ explains the superhorizon growth since the curvature mode will depend on integrals of $\psi$ which does not immediately decay on superhorizon scales.

We can make use of the fact that the equation of motion of $v_\zeta$  is free if we define it as $v_\zeta=\dot{\tilde\zeta}$ instead of the usual definition. In order to do this consistently  we  also  require the  noise of the field $v_\zeta$ to include the superhorizon growth. This can be done by shifting $D_\zeta$ to be
\begin{align}
D_\zeta \to 9H^3\Delta_\zeta^2(t)\equiv9H^4\Delta_\zeta^2 \left(1+36\frac{H^2\Omega^2}{\mu^4}(1-e^{-t/t\psi})^2\right) ,
\end{align}
where $\Delta_\zeta^2(t)$ is the same as that given in (\ref{time_dependent_ps-1}). We defer a more detailed analysis on how to modify the noise term to  Appendix \ref{UL_section}. 
Since the variable that appears in the equations of motion is $v_\zeta$ there is no difference between which variable we use, which implies that the statistics of $\tilde\zeta$ are the same than for $\zeta$. We can check explicitly by solving the Fokker-Planck equation, which now becomes
\begin{align}
\frac{\partial P}{\partial t}=+\frac{2\Delta_\zeta^2(t)\Omega}{H}\frac{\partial^2}{\partial\tilde\zeta\partial\psi}(\psi P)+\frac{H\Delta_\zeta^2(t)}{2}\frac{\partial^2P}{\partial\tilde\zeta^2}+\frac{\partial}{\partial\psi}\left(t_\psi^{-1}\psi P+\frac{D_\psi}{2}\frac{\partial P}{\partial\psi}\right) .
\end{align}
This equation can be solved using the  techniques  described in Appendix \ref{sec:General_solution}. The solution is found to be given by
\begin{align}
P(\zeta,\psi,t)\sim\exp\left(-\frac{\psi^2}{\sigma_\psi^2}-\frac{1}{2\sigma_\zeta^2-2\kappa^2/\sigma_\psi^2}\left(\tilde\zeta-\frac{\kappa}{2\sigma_\psi^2}\psi\right)^2\right) ,
\label{linear_PDF}
\end{align}
where the variances are defined in \eqref{eq:variance_zeta} and \eqref{eq:variance_psi}, and
\be
\kappa\equiv \frac{2}{3} t_\psi \Delta_\zeta^2 f_\zeta \Omega(1-e^{-t/t_\psi}) ,
\label{kappa}
\ee
is the off-diagonal variance which grows from an initial value $\kappa\sim H t \Omega/f_\zeta$ in the ultralight phase to a value  of $\kappa\sim H t_\psi \Omega/f_\zeta$ when the distribution for $\psi$ reaches equilibrium.  Notice that the combination $\kappa/\sigma_\psi^2$ is almost constant and is related to the size of the coupling between the two fields. Also we have that $\sigma_\zeta^2\sigma_\psi^2\gg \kappa^2$ and so the denominator of the second term in the PDF is always positive. 

After marginalising over $\psi$ we find that the variance of $\tilde \zeta$ is given by \eqref{eq:variance_zeta}, as anticipated, hence the  Gaussian statistics of this field are independent on how we define $v_\zeta$. Clearly the same happens for $\psi$. Moreover, since all couplings are linear, the variances do not change but the minima of $\psi$ is displaced. By minimizing the PDF with respect to $\psi$ we find that the minima $\bar\psi$ is at
\begin{align}
\bar \psi=\frac{\kappa}{2\sigma_\zeta^2}\zeta\sim \frac{ \Omega}{H}f_\zeta\tilde\zeta, 
\end{align}
 which implies that the classical trajectory of $\psi$ is shifted by the interaction. Of course, this does not mean that the statistical fluctuations are modified, since they are still simply given by $\sigma^2_\psi$. From now on, for simplicity, we will remove the tilde from $\zeta$.

\section{Non linear interactions in a two-field model}
\label{sec:nonlinear}

So far we have focused our attention on the second order action (\ref{Action-two-field-linear}) where it was possible to find an exact solution of the Fokker-Planck equation. Now we would like to consider the role of non linear terms and study how they modify the probability distribution~\eqref{linear_PDF}. As already stated, for a spectator field there are well known techniques which allow us to find non-perturbative solutions. However these techniques  are not useful to uncover the PDF for the curvature field $\zeta$ since, as we have discussed, there is no equilibrium distribution for $\zeta$. Despite this shortcoming, we will be able to uncover precise non-perturbative effects on the  joint distribution $P(\zeta,\psi)$, based on the Gaussian distribution we found in the previous section. To do so, our strategy will be to ignore non linear terms in $v_\zeta$, while keeping higher order terms in $\psi$. 

Let us start this discussion by writing down the action for perturbations in the case of a canonical two-field model of inflation~\cite{Achucarro:2012sm,Garcia-Saenz:2019njm}. Up to leading order in slow-roll the action is,
\begin{align}
S&=\frac{1}{2}\int d^4 xa^3\left\{\left(f_\zeta+\frac{\Omega}{H}\psi\right)^2\left(\dot\zeta^2-\frac{(\nabla\zeta)^2}{a^2}\right)-2\frac{\Omega}{H}(2f_\zeta H+\Omega \psi)\psi\dot\zeta\right.\nonumber\\
&\hspace{2.3cm}\left.+\dot\psi^2-\frac{(\nabla\psi)^2}{a^2}-V(\psi) \right\}+\dots , \label{two-field-action-non-linear}
\end{align}
For simplicity we consider the first few powers of the potential $V(\psi)=m^2\psi^2+\frac{\lambda}{3}\psi^3+\frac{g}{12}\psi^4 + \cdots$. A crucial point is that the action contains non linear interactions between $\zeta$ and $\psi$, which are due to the non geodesic motion of the background trajectory. Apart from those appearing explicitly in (\ref{two-field-action-non-linear}),  there are no further interactions between the two fields, which can be understood as arising from the original canonical kinetic term in the action. Non canonical kinetic terms will generate higher order interactions between $\dot\zeta$ and $\psi$ which we are assuming to be suppressed. Furthermore, notice that we have not written interactions including gradients coming from gravitational couplings, as they will be negligible for the stochastic dynamics. Finally, as in the linear case, it will be convenient to write the action in terms of $v_\zeta=\dot\zeta-\frac{2\Omega}{f_\zeta} \psi$. Doing this, the action becomes
\begin{align}
S&=\frac{1}{2}\int d^4 xa^3\left\{f_\zeta^2\left(\dot\zeta-\frac{2\Omega}{f_\zeta}\psi\right)^2-f_\zeta^2\frac{(\nabla\zeta)^2}{a^2}+\dot\psi^2-\frac{(\nabla\psi)^2}{a^2}-\mu^2\psi^2\right.\nonumber\\
&\hspace{2.2cm}\left.+\frac{6\Omega^2}{H}\psi^2\left(\dot\zeta-\frac{2\Omega}{f_\zeta}\psi\right)+\frac{2f_\zeta \Omega}{H}\psi\left(\dot\zeta-\frac{2\Omega}{f_\zeta}\psi\right)^2-\frac{\tilde\lambda}{3}\psi^3+ \cdots \right\} \ ,
\label{Lagrangian_v}
\end{align}
where  we have kept terms up to cubic order with respect to $\psi$. Higher interactions are suppressed by further powers of $\Omega/H$ (which we take as a small parameter) although mixed terms are only up to fourth order and they can be reincorporated without trouble.  Notice that in the same way as the mass term of $\psi$ becomes the entropic mass $\mu$, other self interaction couplings are also modified, $\tilde g=g-\frac{12\Omega^3}{f_\zeta H}$ and $\tilde \lambda=\lambda-48 \frac{\Omega^4}{H^2f_\zeta^2}$.

For simplicity, let us examine the case when there is a large cubic  interaction for $\psi$ but the equation for $\zeta$ can be considered as free. The  equations of motion are
\begin{align}
\frac{d}{dt}v_\zeta+3Hv_\zeta+ \frac{k^2}{a^2}\zeta&=-6\frac{\Omega}{f_\zeta}v_\zeta\psi-\frac{6\Omega^2}{Hf_\zeta^2}(3H\psi^2+2\dot\psi \psi)\ , \label{eq:full_EOMa}\\
\ddot\psi+3H\dot\psi+\frac{k^2}{a^2}\psi+\mu^2\psi&=-2\Omega f_\zeta v_\zeta+\frac{2\Omega^2}{H}\psi v_\zeta+\frac{\Omega f_\zeta}{H}v_\zeta^2 \ ,
\label{eq:full_EOM}
\end{align}
where we have disregarded higher order self interactions of $\psi$ (which can be included back at any point of our analysis). In order to apply the stochastic approximation for $\zeta$ we have to demand that higher order interactions are suppressed. The first term on the RHS of Eq.~\ref{eq:full_EOM} is suppressed for typical fluctuations, since $\Omega \ll f_\zeta$. We will assume that the third term is at most of the size  of  the second one. For the second term on the RHS we have that
\begin{align}
    \frac{6\Omega^2}{f_\zeta^2}\frac{\psi^2}{H v_\zeta}\ll 1 \ ,
    \label{eq:interactions_supression}
\end{align}
which follows from the fact that $\Omega\ll f_\zeta$ and that the variance of $\zeta$ is much larger than the one for $\psi$ for $\Omega\neq 0$. For larger fluctuations of  $\zeta$ the inequality \eqref{eq:interactions_supression} still holds. Since at leading order Eq. \eqref{eq:full_EOMa} is free then $v_\zeta$ decays after leaving the horizon and the LHS of Eq.~\eqref{eq:full_EOM} is negligible.

\subsection{Fokker-Planck equation}

In order to study the effect of $v_\zeta$ over $\psi$ more systematically we will analyse the stochastic dynamics of the two-field system. For this, we use the strategy employed for the linear case, that is, we coarse grain the fields directly from the equation of motion for the perturbations \eqref{eq:full_EOM}. Leading non linearities come only from long-wavelength modes, with interactions involving short-wavelength modes being subdominant. In the end, the effects of short wavelength modes reduce to the same linear noise terms as in the linear Langevin equations \cite{Tolley:2008na}.  Moreover, we will consider  the couplings to be small with respect to $H$ so it is possible to treat interactions using perturbation theory. This implies that the noise terms are as in the linear case considered in \eqref{eq:linearisedLangevin}. Another simplification comes from the fact that there are no interactions involving $\dot\psi$ (as they are gravitationally suppressed). Because of this, we can neglect all terms with time derivatives of $\psi$ except for the leading friction term. Indeed this is related to the fact that for typical fluctuations $\ddot\psi\ll H^2\psi$, since in  the long wavelength limit we have that
\begin{align}
\frac{\dot\psi}{H\psi}\sim\frac{2\mu^2}{3H^2}\ll 1 \ ,
\end{align}
where the last inequality follows from the fact that we are  considering light fields. By the same argument we may ignore the second derivative of $\psi$ in the first equation. Of course this can be understood as integrating out $\dot\psi$ from the Fokker-Planck equation and the details will be analogous to those examined in the case of the curvature field in Section~\ref{sec:integrat-v_zeta}. Finally, after separating the equations into long- and short-wavelength modes, we find that the Langevin equations for the long-wavelength mode are
\begin{align}
\frac{dv_\zeta}{dt}+3Hv_\zeta+\frac{6\Omega}{f_\zeta}v_\zeta\psi+\frac{18\Omega^2}{f_\zeta^2}\psi^2&=3H\eta_\zeta , \\
3H \dot\psi+\mu^2\psi+2\Omega f_\zeta v_\zeta-\frac{2\Omega^2}{H}\psi v_\zeta-\frac{\Omega f_\zeta}{H}v_\zeta^2&=\eta_\psi  .
\end{align}
As previously discussed, there are two ways of introducing $\zeta$ to the Langevin equations. We follow the simpler one, whereby we consider an extra Langevin equation for the field $\dot\zeta=v_\zeta$. As we described in Section~\ref{sec:pdf}, this means that we need to include a time dependent noise for $v_\zeta$. After considering these steps, we finally find that the associated Fokker-Planck equation is given by
\begin{align}
\frac{\partial P}{\partial t}=&-\frac{\partial}{\partial\zeta}(v_\zeta P)+\frac{\partial}{\partial v_\zeta}\left(\left(3H v_\zeta+\frac{18\Omega^2}{f_\zeta^2}\psi^2+\frac{6\Omega}{f_\zeta}v_\zeta\psi\right)P\right)  \nonumber\\
&+\frac{\partial}{\partial\psi}\left(\left(\frac{\mu^2}{3H}\psi+\frac{2\Omega f_\zeta}{3H}v_\zeta-\frac{2\Omega^2}{3H^2}\psi v_\zeta-\frac{\Omega f_\zeta}{3H^2}v_\zeta^2\right)P\right)
+\frac{D_\psi}{2}\frac{\partial^2}{\partial \psi^2}P+\frac{9}{2} H^3\Delta_\zeta^2 (t)\frac{\partial^2P}{\partial v_\zeta^2} , \label{eq:FPNonGaussian1}
\end{align}
where $\Delta_\zeta^2(t)$ is the same quantity found in Eq. (\ref{time_dependent_ps-1}), $D_\psi=H^3/4\pi^2$, and where the variances are given by
\begin{align}
\sigma_\zeta^2 (t)&=\Delta_\zeta^2 Ht\left(1+\frac{36 H^2\Omega^2}{\mu^2}\right)-54\Delta_\zeta^2\frac{H^4\Omega^2}{\mu^6}\left(3-4e^{-t/t_\psi}+e^{-2t/t_\psi}\right)  , \\
\sigma_\psi^2 (t)&=\frac{3H^4}{8\pi^2\mu^2}\left(1-e^{-2t/t_\psi}\right) .
\end{align}
Before continuing, let us comment on the fact that drift terms including powers of $v_\zeta$ will become $\zeta$ derivatives. This can be understood as a consequence  of the shift symmetry of the curvature mode (see also \cite{Cohen:2021fzf}). Using this we can deduce that terms including two derivatives of $\zeta$ will change the variance of the $\zeta$ distribution, and the tail of the distribution of $\psi$. Since we are interested in the tail of the distribution of $\zeta$ we can ignore them for now and include them later.  This is achieved by imposing that the quadratic terms in the drift for $\psi$ is larger than the quadratic terms in the drift in $v_\zeta$ or, equivalently, that
\begin{align}
\frac{\Omega}{H}\frac{f_\zeta \zeta}{\psi}\gg 1 \ ,
\label{Neglecting_vz_drift}
\end{align}
where we have used the fact that $v_\zeta\sim H\zeta$.  Notice that this is achieved only for $\Omega $ relatively large, although not necessarily larger than $H$. A non zero  $\Omega$ increases the  variance of $\zeta$ making it much larger than that of $\psi$, which otherwise will be very similar.  Besides that, if $\Omega/H$ is very suppressed, it  will make the inequality in \eqref{Neglecting_vz_drift} impracticable.
In what follows we will assume that \eqref{Neglecting_vz_drift} holds, and we will comment on its  effect on the PDF later.
Similarly as we saw in Section~\ref{sec:integrat-v_zeta}, terms in the drift containing powers of $v_\zeta$ transform into derivatives of $\zeta$, $v_\zeta\to \Delta_\zeta^2\frac{\partial}{\partial_\zeta}$, which we will show is due to the shift symmetry of $\zeta$. Using this property,  we  can neglect  the term proportional to $v_\zeta^2$ in the drift of $\psi$ since is subleading with respect to  the term proportional to $\psi v_\zeta$.

\subsubsection{Adiabatic elimination of $v_\zeta$}
\label{Adiabatic_elimination}

We have used the variable $v_\zeta$  since it was a convenient way of studying the derivative couplings that appear in the action. Nevertheless, we are not interested in the statistical properties of $v_\zeta$ and moreover it decays faster than the other fields. It is then useful to eliminate $v_\zeta$ from the Fokker-Planck equation and obtain $P(\zeta,\sigma)$ directly. The way of doing this systematically is called adiabatic elimination of fast variables~\cite{VanKampen:1985,van1992stochastic}. We made a similar computation in Section~\ref{sec:integrat-v_zeta}, which we now generalise to include the coupling with another field. The idea is to expand the probability density function in powers of the time  scale of the fast variable. Then replacing order by order it will be possible to factorise the  dependence on the fast variable from the slow variables. For our case, it is convenient to write the Fokker-Planck equation as
\bea
P(\zeta,v_\zeta,\psi;t) &=& P^{(0)}(\zeta,v_\zeta,\psi;t)+(Ht_v) P^{(1)}(\zeta,v_\zeta,\psi;t) \nonumber \\ 
&& +(Ht_v)^2P^{(2)}(\zeta,v_\zeta,\psi;t)+ \cdots .
\label{P_expansion}
\eea
This expansion becomes useful if we write the Fokker-Planck equation as
\begin{align}
\frac{d}{dv_\zeta}\left(v_\zeta+\frac{3}{2}H^2 \Delta^2_\zeta(t)\frac{d}{dv_\zeta}\right)P =\qquad \qquad \qquad \qquad \qquad \qquad \qquad \qquad \nonumber\\t_v\left(\frac{d}{dt}+v_\zeta\frac{d}{d\zeta}-\frac{\partial}{\partial \psi}(t_\psi^{-1} \psi+\frac{2\Omega f_\zeta}{3H}v_\zeta-\frac{2\Omega^2}{3H^2}\psi v_\zeta)+\frac{D_\psi}{2}\frac{\partial^2}{\partial\psi^2}\right)P  \ ,
\label{FP_time_scales}
\end{align}
where we have written the mass in terms of the time scale $t_\psi^{-1}=\mu^2/3H$. Notice that the operator on the RHS of Eq.~\eqref{FP_time_scales} is of order $t_v/t_\psi\ll 1$ whereas the operator on the LHS is of order one,  which is the reason why the approximation is well justified. After replacing \eqref{P_expansion} into the Fokker-Planck equation this can also be written as a series in powers of $t_v$.  At zeroth order in $t_v$ we find the following equation
\begin{align}
\frac{\partial}{\partial v_\zeta}(v_\zeta P^{(0)})+\frac{3}{2}H^2\Delta^2_\zeta(t)\frac{\partial^2P^{(0)}}{\partial v_\zeta^2}=0   .
\label{eq:P0}
\end{align}
Notice that this equation only specifies the  $v_\zeta$ dependence of $P^{(0)}$.  
Since this is a first order ODE we can solve it and write $P^{(0)}$ as
\begin{align}
P^{(0)}(\zeta,v_\zeta,\psi;t)=e^{-\frac{v_\zeta^2}{3H^2\Delta_\zeta^2(t)}}\phi_0(\zeta,\psi;t)
\label{P_0} \ ,
\end{align}
which is a  Gaussian distribution of $v_\zeta$ with variance given by $(3/2) H^2\Delta_\zeta^2(t)$ in line with the linear solution for $P$. Of course, had we included higher order terms in $v_\zeta$, they  would have entered into the LHS of \eqref{FP_time_scales}, and the solution \eqref{P_0} would have had to be modified accordingly.
To figure out what the restrictions are for $P^{(1)}$ we  plug \eqref{P_0} into the expansion for $P$. We find that at order $t_v$
\begin{align}
&\frac{\partial}{\partial v_\zeta}\left(v_\zeta +\frac{3}{2}H^2\Delta^2_\zeta(t)\frac{\partial}{\partial v_\zeta}\right)HP^{(1)}=\left(\frac{\partial}{\partial\psi}(t_\psi^{-1}\psi\phi_0)+\frac{D_\psi}{2}\frac{\partial^2\phi_0}{\partial\psi^2}-\frac{\partial\phi_0}{\partial t}+\right.\nonumber\\
&\qquad\left.+\left(\frac{2\Omega f_\zeta}{3H}\frac{\partial \phi_0}{\partial\psi}-\frac{2\Omega^2}{3H^2}\frac{\partial}{\partial\psi}(\psi\phi_0)-\frac{\partial\phi_0}{\partial \zeta}\right)v-\frac{\dot\Delta_\zeta^2(t)}{3H^2\Delta_\zeta^4(t)}\phi_0 v_\zeta^2\right)e^{-\frac{v_\zeta^2}{3H^2\Delta_\zeta^2(t)}} \ .
\label{gamma1}
\end{align}
Which has a similar structure to \eqref{eq:P0} but with a more complicated RHS. Due to this, it is not possible to immediately solve for $P^{(1)}$. Yet, it is possible to simplify the last  equation by noticing that the LHS is a total derivative and that the RHS is multiplied by a Gaussian function. After integrating over $v_\zeta$ the LHS vanishes while on the RHS only some terms with even powers of $v_\zeta$ remain. Since these have to add up to zero, we finally find
\begin{align}
\frac{\partial}{\partial\psi}(t_\psi^{-1}\psi\phi_0)+\frac{D_\psi}{2}\frac{\partial^2\phi_0}{\partial\psi^2}-\frac{\partial\phi_0}{\partial t}-\frac{\dot\Delta_\zeta^2(t)}{2\Delta_\zeta^2(t)}\phi_0=0  .
\label{Integr_condition1}
\end{align}
Notice that the first three terms form a Fokker-Planck equation for $\phi_0(\psi,t)$ and hence are of order $1/t_\psi$. The last term is of order  $t_v^{-1}$ so we can neglect it. We can in principle  solve for the $\psi$ dependence of $\phi_0$, which is just given by a Gaussian with variance $\sigma_\psi^2$, although the explicit solution will not be important. Instead, we can now find the solution for $P_1$ by using the above condition.  Equation~\eqref{gamma1} then reduces to 
\begin{align}
\frac{\partial}{\partial v_\zeta}\left(v_\zeta +\frac{3}{2}H^2\Delta^2_\zeta(t)\frac{\partial}{\partial v_\zeta}\right)HP^{(1)}=-\left(-\frac{2\Omega f_\zeta}{3H}\frac{\partial\phi_0}{\partial\psi}+\frac{2\Omega^2}{3H^2}\frac{\partial}{\partial\psi}(\psi\phi_0)+\frac{\partial\phi_0}{\partial \zeta}\right)v_\zeta e^{-\frac{v_\zeta^2}{3H^2\Delta_\zeta^2(t)}}  ,
\end{align}
whose solution can be written as
\begin{align}
P^{(1)}=-\frac{v_\zeta}{H}\left(-\frac{2\Omega f_\zeta}{3H}\frac{\partial\phi_0}{\partial\psi}+\frac{2\Omega^2}{3H^2}\frac{\partial}{\partial_\psi}(\psi\phi_0)+\frac{\partial\phi_0}{\partial \zeta}\right)e^{-\frac{v_\zeta^2}{3H^2\Delta_\zeta^2(t)}}+\frac{\phi_1(\zeta,\psi,t)}{H}e^{-\frac{v_\zeta^2}{3H^2\Delta_\zeta^2(t)}} \ .
\end{align}
If we now plug this solution into the terms of order $t_v^2$, we find that
\bea
&& \frac{\partial}{\partial v_\zeta}\left(v_\zeta +\frac{3}{2}H^2\Delta^2_\zeta(t)\frac{\partial}{\partial v_\zeta}\right)H^2P^{(2)} = \Bigg[ \frac{\partial}{\partial\psi}(t_\psi^{-1}\psi\phi_1)+\frac{D_\psi}{2}\frac{\partial^2\phi_1}{\partial\psi^2} \nonumber\\ 
&& 
\qquad 
-\frac{\partial\phi_1}{\partial t}+(\dots)v_\zeta+(\dots)v_\zeta^3 + \bigg( -\frac{4\Omega f_\zeta}{3H}\frac{\partial^2\phi_0}{\partial\psi\partial\zeta}+\frac{4\Omega^2}{3H^2}\frac{\partial^2}{\partial\psi\partial\zeta}(\psi\phi_0) \nonumber\\ 
&& 
\qquad 
+\frac{\partial^2\phi_0}{\partial\zeta^2}-\frac{4\Omega^4}{H^4}(\phi_0+\psi(3\frac{\partial\phi_0}{\partial\psi} +\psi{\frac{\partial \phi_0}{\partial\psi}}) \bigg) v_\zeta^2 \Bigg] e^{-\frac{v_\zeta^2}{3H^2\Delta_\zeta^2(t)}} \ .
\eea
This equation is similar to the constraint for $P^{(1)}$, thus  we can  again integrate over $v_\zeta$. We find that
\bea
&& \frac{\partial}{\partial\psi}(t_\psi^{-1}\psi\phi_1)+\frac{D_\psi}{2}\frac{\partial^2\phi_1}{\partial\psi^2}-\frac{\partial\phi_1}{\partial t}-2\Omega f_\zeta\Delta_\zeta^2\frac{\partial^2 \phi_0}{\partial\psi\partial\zeta} \nonumber \\
&& 
\qquad \qquad +2\Omega^2\Delta_\zeta^2(t)\frac{\partial^2}{\partial\psi\partial\zeta}(\psi\phi_0) +\frac{3H^2\Delta_\zeta^2(t)}{2}\frac{\partial^2\phi_0}{\partial\zeta^2}=0  ,
\eea
where we have neglected sub leading terms. Notice that the terms proportional to $\Omega^4$ can be recast as subleading corrections to the noise of $\zeta$ of the form $\Omega^2\zeta^2$. These terms have been recently studied~\cite{Gorbenko:2019rza,Mirbabayi:2020vyt,Cohen:2021fzf} and we will leave their analysis to future work. Instead of finding a solution for  $P_2$, let us collect the terms in the PDF up to order $t_v$
\begin{align}
P(\zeta,v_\zeta,\psi;t)=\left(\phi_0(\zeta,\psi;t)+t_v\left(v_\zeta\left(\frac{2\Omega^2}{H^2}\frac{\partial}{\partial_\psi}(\psi\phi_0)+\frac{\partial\phi_0}{\partial \zeta}\right)+\phi_1(\zeta,\psi;t)\right)\right)e^{-\frac{v_\zeta^2}{3H^2\Delta_\zeta^2(t)}} \ .
\label{Integr_condition2}
\end{align}
Since we are not interested in the full distribution, but only on the one on configuration space, we can integrate out  $v_\zeta$, which leads to $P(\zeta,\psi,t)=\sqrt{3\pi H^2\Delta_\zeta^2(t)}(\phi_0+t_v\phi_1)$. Finally collecting all the ingredients, taking the time derivative  of $P(\zeta,\psi,t)$ and using \eqref{Integr_condition1} and \eqref{Integr_condition2} we find that
\begin{align}
\boxed{
\frac{\partial P}{\partial t}=\frac{\partial}{\partial\psi}\left( \frac{\psi P}{t_\psi}+\frac{D_\psi}{2}\frac{\partial P}{\partial\psi}\right)+ \frac{2 \Omega}{3} \Delta_\zeta^2(t)\frac{\partial^2}{\partial\psi\partial\zeta}\left(\frac{\Omega}{H}\psi P - f_\zeta P \right)+\frac{H\Delta_\zeta^2(t)}{2}\frac{\partial^2P}{\partial^2\zeta}+\mathcal{O}(t_v) , } 
\label{FP_zetapsi}
\end{align}
which is a  Fokker-Planck equation for $\zeta$ and $\psi$ after integrating out $v_\zeta$. Notice that the couplings between $v_\zeta$ and $\psi$ now translate into mixed derivatives between $\psi$ and $\zeta$. Of course in the absence of such couplings, the system reduces to two linear Fokker-Planck equations. We can find an analytical solution if we Fourier transform $\zeta$ and $\psi$ to $p$ and $q$. The resulting equation can be solved by looking for solutions of the form
\begin{align}
P(p,k;t)=e^{-\frac{\sigma_\zeta^2(t)}{2}p^2}e^{-M(p,t)k^2-L(p,t)k} ,
\end{align}
which translates into two independent ODEs for $M$ and $L$:
\begin{align}
\frac{d M(t)}{dt}+\frac{2}{3H^2t_\psi}(3H^2+2i p t_\psi\Omega^2 H\Delta_\zeta^2(t))M(t)-\frac{D_\psi}{2}=0 ,
\label{M_eq} \\
\frac{d L(t)}{dt}+\frac{1}{3H^2 t_\psi}(3H^2+2i p t_\psi\Omega^2 H\Delta_\zeta^2(t))L(t)-\frac{2 f_\zeta \Omega \Delta_\zeta^2(t)}{3}p=0 .
\end{align}
Imposing that at $t=0$, $M=0$,  we find 
\begin{align}
M(t)=\frac{1}{2}D_\psi e^{-\frac{2t}{t_\psi}-\frac{4ip\Omega^2\sigma_\zeta^2(t)}{3H^2}}\int_0^tdt'e^{\frac{2t}{t_\psi}+\frac{4ip\Omega^2\sigma_\zeta^2(t)}{3H^2}} \ .
\label{M_int}
\end{align}
We can approximate the integral using a saddle point approximation to find
\begin{align}
M(t)&=\frac{3D_\psi  t_\psi}{4(3H+2ipt_\psi\Omega^2 \Delta_\zeta^2(0))}e^{-\frac{2t}{t_\psi}-\frac{4ip\Omega^2\sigma_\zeta^2(t)}{3H^2}}\left(-1+e^{\frac{2t}{t_\psi}+\frac{4ip t\Omega^2\Delta_\zeta^2(0)}{3H}}\right) \ .
\label{M}
\end{align}
This approximation works well for early times $1\ll Ht\ll Ht_\psi$ times, when $\psi$ has not reached its equilibrium distribution. At later times a better approximation is obtained by considering that $M(t)$ is time independent and we will comment on this later. Following the same method a straightforward computation shows that $L(t)$ is
\begin{align}
    L(t)=-\frac{i 2f_\zeta H t_\psi p}{(3H+2ipt_\psi\Omega^2 \Delta_\zeta^2(0))}e^{-\frac{t}{t_\psi}-\frac{2ip\Omega^2\sigma_\zeta^2}{3H^2}}\left(1-e^{\frac{t}{t_\psi}+\frac{2ip\Omega^2 \Delta_\zeta^2(0)}{3H}}\right).
    \label{L}
\end{align}
Using the solution \eqref{M} and \eqref{L} we can immediately compute the Fourier transform in $k$ since  the PDF is Gaussian. Computing the Fourier transform in $p$ is non trivial, since $p$ appears in the exponents. In order to Fourier transform we expand  $M(t)$ and $L(t)$ in powers of $\Delta_\zeta^2(0)$. Notice that this requires $\zeta\gg \Delta_\zeta^2(0) \Omega^2/\mu^2$ which we will assume to hold true and indeed it does for typical values of $\zeta$ and small amplitude of the density perturbations. Retaining terms only  up to second order in $p$ and Fourier transforming back, we finally obtain:
\begin{align}
\boxed{
P(\zeta,\psi)=\exp\left[ -\frac{\psi^2}{2\sigma_\psi^2}-\frac{1}{2\sigma_\zeta^2}\left(\zeta+\frac{2f_\zeta \Omega}{3H}\frac{\sigma_\zeta^2}{\sigma_\psi^2}\psi-\frac{\Omega^2}{3H^2}\frac{\sigma_\zeta^2}{\sigma_\psi^2}\psi^2\right)^2\right] . }
\label{Joint_distribution_Complete}
\end{align}
which is valid for $1\ll Ht\ll Ht_\psi$.

\subsection{Non Gaussian tails}

In order to understand the effect of the non linear interactions in the PDF it is useful to ignore first the linear mixing term. This is not well justified since it means that we are ignoring a  $\psi \zeta$ interaction that is important for large $\zeta $. Nevertheless its addition does not change the qualitative effect of adding the  non linear  derivative interaction. As we discuss in Section~\ref{sec:pdf}, this is because the effect of the linear mixing term is shifting the trajectory of $\psi$ but not its variance. Due to this let us ignore this term for the moment. Furthermore, we see that by doing this, it becomes possible to integrate over $\psi$ analytically which greatly simplifies the analysis. After these considerations let us study the following  PDF
\begin{align}
\boxed{
P(\zeta,\psi)=\exp\left[ -\frac{\psi^2}{2\sigma_\psi^2}-\frac{1}{2\sigma_\zeta^2}\left(\zeta-\bar\kappa\frac{\psi^2}{2\sigma_\psi^2}\right)^2\right],}
\label{Joint_distribution}
\end{align}
where we have defined 
\begin{align}
\bar\kappa\equiv \frac{\Omega^2}{3H^2}\sigma_\zeta^2,
\end{align}
which is a time independent parameter. Notice that this coupling is related to \eqref{kappa} through the relation
\begin{align}
\bar \kappa=\frac{\kappa^2}{\sigma_\psi^2},
\end{align} 
by which it should be clear that $\kappa$ is related to the size of the interactions. We are interested in the distribution for $\zeta$ which we obtain after integrating over $\psi$. The integral can be done analytically and expressed in term of Bessel functions:
\begin{align}
P(\zeta)\propto
\begin{cases}
\exp\left(\frac{(\zeta\bar\kappa-\sigma_\zeta^2)^2}{4\bar\kappa^2\sigma_\zeta^2}-\frac{\zeta^2}{2\sigma_\zeta^2}\right)K_{1/4}\left(\frac{(\zeta\bar\kappa-\sigma_\zeta^2)^2}{4\bar\kappa^2\sigma_\zeta^2}\right)\hspace{4.5cm}\mathrm{for}\ \zeta<\frac{\sigma^2_\zeta}{\bar\kappa} \\
\exp\left(\frac{(\zeta\bar\kappa-\sigma_\zeta^2)^2}{4\bar\kappa^2\sigma_\zeta^2}-\frac{\zeta^2}{4\sigma_\zeta^2}\right)\left(I_{-1/4}\left(\frac{(\zeta\bar\kappa-\sigma_\zeta^2)^2}{4\bar\kappa^2\sigma_\zeta^2}\right)+I_{1/4}\left(\frac{(\zeta\bar\kappa-\sigma_\zeta^2)^2}{4\bar\kappa^2\sigma_\zeta^2}\right)\right)\qquad \mathrm{for}\ \zeta>\frac{\sigma^2_\zeta}{\bar\kappa}
\end{cases} \!\!\! .
\label{cases:NED}
\end{align}
\begin{figure}

\includegraphics[scale=0.45,trim=1.8cm 8cm 1cm 4cm,clip]{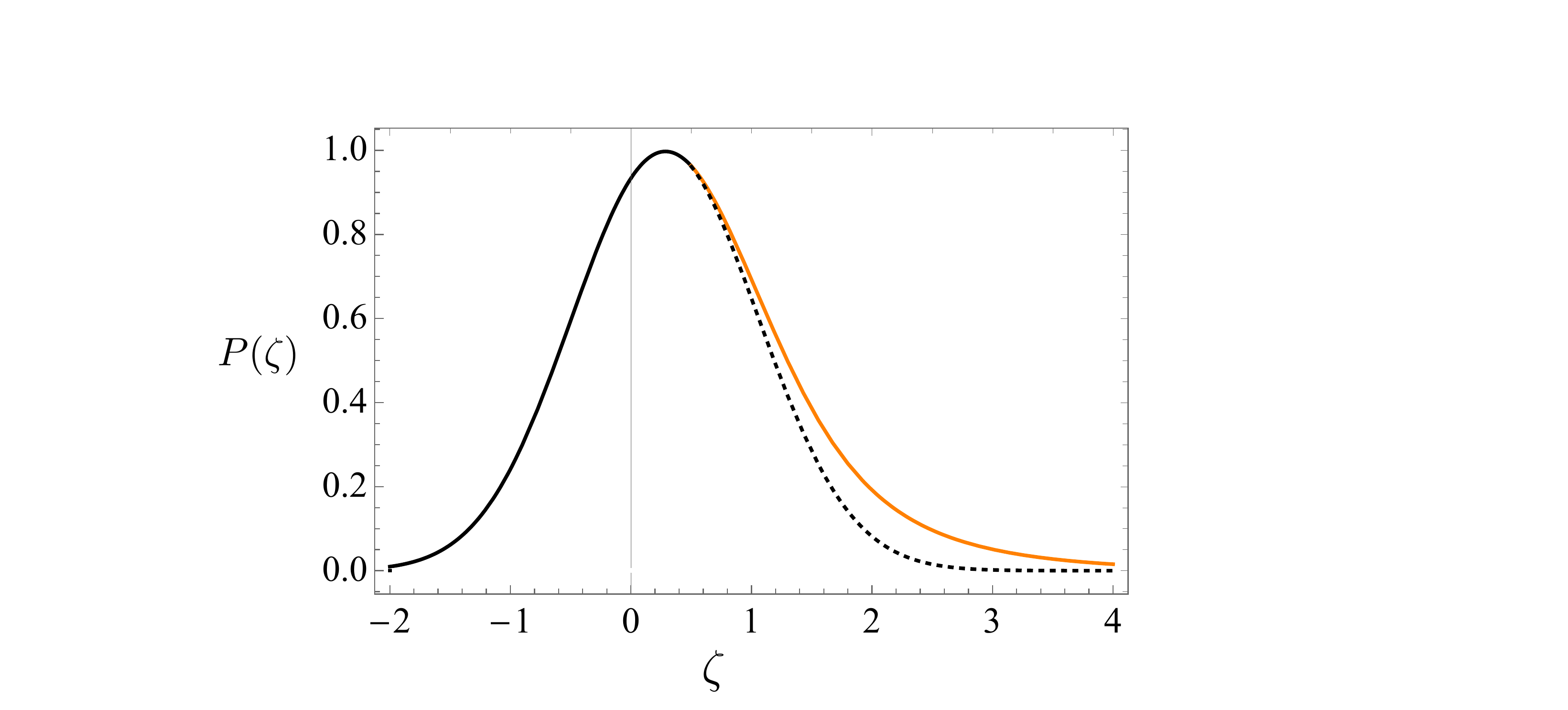}
\caption{Toy model of the distribution described in \eqref{cases:NED}. The black region is the when $ \zeta< \sigma_\zeta^2/\bar\kappa$ while the orange line is otherwise. The dotted line is a Gaussian distribution that fits the region in the left of the plot. The parameters are made up to highlight the fact that the distribution on the right  has a very non Gaussian tail.}
\label{Fig:distribution_A}
\end{figure}
To appreciate the distribution we plot it in Fig.~\ref{Fig:distribution_A}. From there we see that for it is a displaced Gaussian  around the centre but it becomes strongly non Gaussian for $\zeta>\zeta_{\mathrm{cr}}\equiv\frac{\sigma^2_\zeta}{\bar\kappa} $.
We can understand the asymptotic behavior of  $P(\zeta)$  as emerging from a change on the saddle points in  \eqref{Joint_distribution}.  Let us notice from it,  that the shifted distribution of $\zeta$ has the overall effect of changing the coefficient in front of $\psi^2$, which becomes negative for large values of $\zeta>\zeta_{\mathrm{cr}}$. This implies that there are  three saddle points for $\psi$, one at $\psi=0$, and other two at
\begin{align}
\bar\psi=\pm\sqrt{\frac{2\sigma_\psi^2}{\bar\kappa}}\sqrt{\zeta-\frac{\sigma_\zeta^2}{\bar\kappa}} \ .
\end{align}
The behaviour for the asymptotics of $P(\zeta)$ are then similar to the Stokes phenomena, in the sense that for large values of $\zeta$ the saddle point changes from $0$ to $\bar\psi$. If we expand around $0$  we find a Gaussian distribution for $\zeta$, whereas if we expand \eqref{Joint_distribution} around $\bar\psi$ after integrating over $\psi$, we find that
\begin{align}
P(\zeta)\sim\exp\left(-\frac{\zeta}{\bar\kappa}\right),
\end{align}
valid for $\zeta\gg \zeta_{\mathrm{cr}}$ and that it coincides with the large $\zeta$ limit of \eqref{cases:NED}. This effect is non perturbative in nature since   $\bar\kappa\ll 1$ and so typically $\zeta/\bar\kappa\geq 1$. Let us now comment on the regime of validity of $\bar\kappa$. As we mentioned  before Eq. \eqref{M} was only valid for intermediate times $t_v\ll t\ll t_\psi$, whereas for  later $ t\geq t_\psi$ it is more accurate to consider that $M(t)$ is time independent. Solving for $M$ and $L$ we find that the PDF is given by,
\begin{align}
P(\zeta,\psi)=\exp\left(-\frac{\psi^2}{2\sigma_{0\psi}^2}-\frac{1}{2\sigma_\zeta^2}\left(\zeta-\bar\kappa'\frac{\psi^2}{2\sigma_{0\psi}^2 }\right)^2\right)\ ,\qquad
\bar\kappa'=\frac{1}{3} t_\psi H\Delta_\zeta^2(t)\frac{\Omega^2}{H^2} \ ,
\label{Joint_distribution_equilibrium}
\end{align}
with $\bar\kappa$  replaced by $\bar\kappa'$ and $\sigma_{0\psi}^2=3H^4/(8\pi^2\mu^2)$ the equilibrium distribution for $\psi$. In absence of $\zeta$, the PDF in \eqref{Joint_distribution_equilibrium} reduces to the equilibrium distribution of $\psi$, in accordance with the fact the we are considering the distribution at late times $t\geq t_\psi$. In this sense, \eqref{Joint_distribution_equilibrium} is the distribution when the field $\psi$ has settled into its equilibrium distribution. Of course if the field is ultralight then $t_\psi\to\infty$ and the transition between $\bar\kappa$ and $\bar\kappa'$ does not take place during inflation.

Furthermore, we see that  for $t\ll t_\psi$, $\bar\kappa<\bar\kappa'$ but otherwise $\bar\kappa>\bar\kappa'$ since $\bar\kappa$ keeps growing. Expanding in powers of $t_\psi$ we find that
\begin{align}
\bar\kappa=\frac{\Omega^2}{3H^2}Ht\Delta_\zeta^2(t) \ ,\qquad \bar\kappa'=\frac{\Omega^2}{3H^2}Ht_\psi\Delta_\zeta^2(t) \ ,
\end{align} 
from where we can see that $\bar\kappa$ grows until it reaches the equilibrium value $\bar\kappa'$. Anyhow since at equilibrium $\sigma_\zeta\approx Ht_\psi\Delta_\zeta^2$, we can write $\kappa'\sim \Omega^2/H^2 \sigma_\zeta^2$ to see that the coupling does not change its dependence on the parameters. In any case is worth mentioning that $\bar\kappa'$  is a limiting value where the time dependence has become negligible.

\begin{figure}
  \includegraphics[scale=0.48,trim=0.5cm 13cm -1cm 3cm,clip]{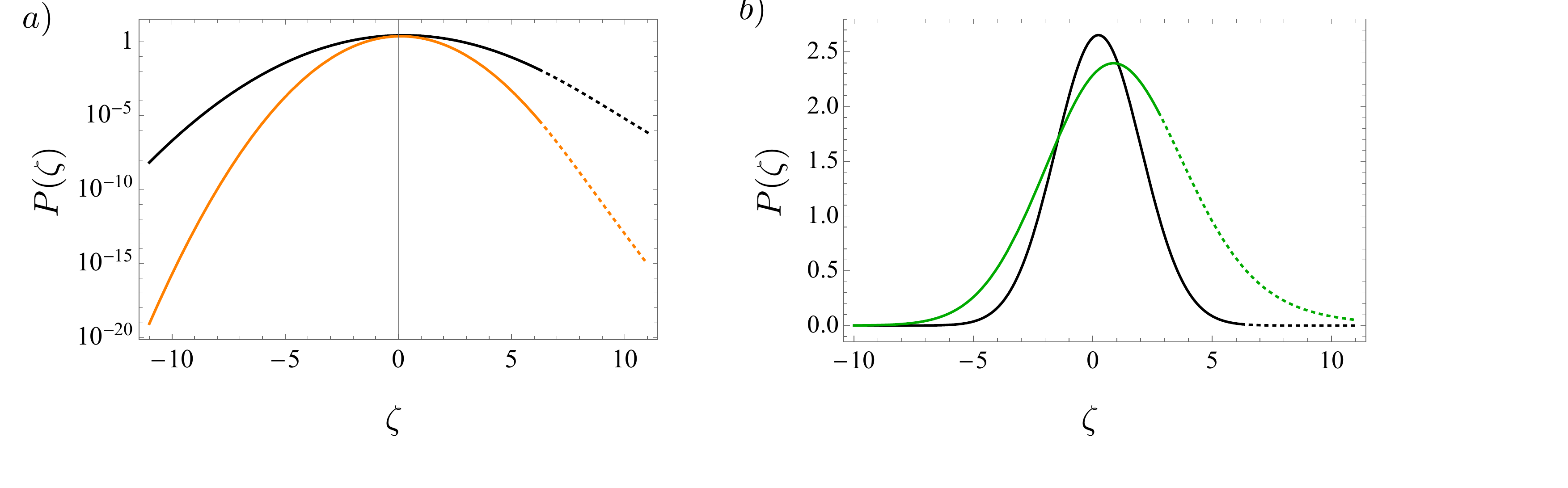}
\caption{a) The figure is the PDF of \eqref{cases:NED}. The orange line is for $N=30$ and the black line is for $N=40$. The other parameters are $\Omega=.2H$ and $\Delta_\zeta=10{-3}$. The dotted lines are the region where $\zeta\geq \sigma_\zeta^2/\bar\kappa$ b) Both curves are at $N=40$ but the green line has $\Omega=0.3H$. }
\label{Fig:tails_ch_par}
\end{figure}

All of this implies that initially the tail of the distribution becomes non Gaussian at smaller values of $\zeta$ until it settles down on $\sigma_\zeta^2/\bar\kappa'$. When this happens the coupling of the tail becomes constant
\begin{align}
P(\zeta)\sim\exp\left(-\frac{\zeta}{\bar\kappa}\right)=\exp\left(-\frac{\mu^2}{2\Omega^2\Delta_\zeta^2(t)}\zeta\right).
\end{align} 
Still at early times the distribution is more localised, hence the values at which the tail becomes non-Gaussian are smaller, as can be seen in Fig.~\ref{Fig:tails_ch_par}. In the end, even though the saddle point changes at larger $\zeta$, larger values of $\zeta$ are more likely due to the growth of $\sigma_\zeta^2$.  Finally, let us pay attention to the fact that  the tail  is typically very suppressed since we have that
\begin{align}
\zeta/\bar\kappa\gg \sigma_\zeta^2/\bar\kappa^2\gg 1 \ ,
\end{align} since $
\bar\kappa^{-1}\gg 1$. In order to have a larger effect the power spectrum has to be several orders of magnitude larger than the CMB values. Furthermore, let us note that the critical value for $\zeta$ does not depend on the amplitude of the fluctuations $\Delta_\zeta^2(t)$, which in the end implies that for smaller values of $\Delta_\zeta^2(t)$ the probability for the regions where the tail changes is very suppressed. In any case the tail we have found is always larger than a Gaussian tail, which is due to the fact that $\bar\kappa\zeta/\sigma_\zeta^2>1$ is equivalent to $\zeta/\bar\kappa>\zeta^2/\sigma_\zeta^2$.

To conclude let us stress the point that changing $\Omega$ has a large effect, as can be seen in Fig.\ \ref{Fig:tails_ch_par}b, where the overall effect is flattening and shifting the PDF.

\subsection*{Local non-Gaussianity}

In order to understand how the tail is related to usual perturbation theory, we can estimate the size of the non-Gaussianities produced by the interaction $\sim v_\zeta^2\psi$ we are considering.  From the Langrangian \eqref{Lagrangian_v}, we have that $f_{\mathrm{NL}}$ can be estimated to be of order 
\begin{align}
f_{\mathrm{NL}}\zeta\sim \frac{\mathcal{L}_3}{\mathcal{L}_2}\sim\frac{6\Omega^2}{H^2 f_\zeta^2}\frac{\psi^2}{\zeta}\sim\frac{\Omega^2}{H^2 f_\zeta^2}\frac{\sigma_\psi^2}{\sigma_\zeta^2}\zeta \ ,
\end{align}
where the last term corresponds to the off diagonal coupling parameter. Indeed, using \eqref{kappa} the last relation can be recast as
\begin{align}
f_{\mathrm{NL}}\zeta\sim \frac{\Omega^2}{H^2}\zeta \ ,
\end{align}
which implies the following: Perturbation theory is usually under control when $f_{\mathrm{NL}}\zeta\ll 1$ which in our case corresponds to expanding around the Gaussian saddle point. Observables can be computed by expanding them in powers of the power spectrum since it is always suppressed.  On the other hand, when $f_{\mathrm{NL}}\zeta\sim 1$ perturbation theory fails since the distribution is non Gaussian, which means that the expansion in powers of the power spectrum cannot be justified. We can see this explicitly now since $\zeta_{\mathrm{cr}}\sim1/f_{\mathrm{NL}}$. Naively one should expect that for $\zeta\geq \zeta_{\mathrm{cr}}$ perturbation theory fails which we now see translates into the tail becoming strongly non-Gaussian. Clearly if $f_\mathrm{{NL}}\ll1$ one might be worried about corrections from other interactions becoming important for large values of the tail. Nevertheless, since there is a finite number of interactions in the EFT we are considering we can still deduce the behaviour of the tail up to certain   (large) values of $\zeta$ when other interactions become important. In order to estimate the validity of our results we need to include the terms we have been neglecting so far, which is what we are going to discuss now.

\subsubsection*{Including the linear mixing}

In the previous section we ignored the linear mixing term, since we argued it does not affect the appearance of non-Gaussian tails. When adding it, it turns out that it is not possible to integrate over $\psi$ analytically but it is still possible to obtain the saddle points and check how the tail changes for large $\zeta$. Firstly, let us notice that the point where the saddle point changes is not modified significantly. To see this we can  expand the PDF and notice that the terms proportional to $\psi^2$ in Eq.~\eqref{Joint_distribution_Complete} are
\begin{align}
\frac{\psi^2}{2\sigma_\zeta^2}\left(1-\frac{4\Omega^2}{3H^2}\left(\zeta-\frac{1}{3}\right)\right),
\end{align}
which implies that for small $\zeta<1/3$ the saddle point does not change even when $\Omega^2/H^2$ is very large. Since we are interested in small $\Omega^2/H^2$ this does not significantly change our results. Next, if we ignore suppressed terms we find that the displaced saddle points are now at
\begin{align}
\bar\psi=\frac{3 f_\zeta H}{\Omega}\pm \frac{3 f_\zeta H^2}{\Omega^2}\sqrt{-1/2+(1+24\zeta)\frac{\Omega^2}{36 H^2}}.
\end{align}
The fact that the saddle point now contains a constant piece is reflected in the tail. Indeed, expanding around $\bar\psi$ we find that for large $\zeta$
\begin{align}
P(\zeta)\sim\exp\left(-\frac{3\zeta\pm\sqrt{6\zeta}}{4\sigma_\zeta^2}\frac{H^2}{\Omega^2}\right),
\end{align}
where the different signs correspond to different saddle points. In general there is one correct saddle point which  can be picked based on the analytical properties of the PDF, a task which we leave for future work.\footnote{See Refs.~ \cite{Witten:2010cx,Feldbrugge:2017kzv,Serone:2017nmd} for related attempts in dealing with complex saddle points.} In  any case, if we assume that the fluctuation in $\zeta$ is large we find that the tail gets a small correction. Finally let us notice that the effects of the linear mixing term can also be understood as $f_{\mathrm{NL}}\zeta$ becoming large. In this case it corresponds to the interaction $\mathcal{L}_3\sim 2 f_\zeta\Omega/H \psi v_\zeta^2$, and so the constant piece of the saddle point, which appears for smaller $\zeta$, appears as long as $\Omega/H\zeta\sim 1$. Nevertheless, the tail of the distribution will not change until $ \Omega^2/H^2\zeta\sim 1$ where the other saddle point becomes real.

As a side comment  let us notice that the coupling does not grow unbounded but it settles into a constant value at approximately $t\sim t_\psi$, given by
\begin{align}
\frac{2f_\zeta\Omega}{3H^2}\frac{t_\psi \Delta_\zeta^2(t)}{\sigma_\psi^2}\psi\sim \frac{\Omega t_\psi}{f_\zeta } \ ,
\end{align}
as explained in Eq.~\eqref{kappa}.

\subsection{Adding more interactions}

Having understood the leading interaction effect we can now add the rest of the terms to the system. In particular, let us consider a general potential $V(\psi)$.
Since the new terms do not add any new shorter time scale than $t_v$ we can eliminate $v_\zeta$ from the Fokker-Plank equation following the same steps we  described in Section~\ref{Adiabatic_elimination}. This results in the following Fokker-Plank equation
\begin{align}
\frac{\partial P}{\partial t}&=\frac{\partial}{\partial\psi}\left(\frac{V'(\psi)}{3H} P+\frac{D_\psi}{2}\frac{\partial P}{\partial\psi}\right)+H\Delta_\zeta^2(t)\frac{\partial^2}{\partial\psi\partial\zeta}\left(\left(\frac{2\Omega^2}{H^2}\psi +\frac{2 f_\zeta\Omega}{3H} \right)P\right)\nonumber\\
&+\frac{H\Delta_\zeta(t)^2}{2}\frac{\partial^2P}{\partial \zeta^2}+\frac{6\Omega^2}{f_\zeta^2 H}\frac{\partial}{\partial \zeta}\left(\psi^2 P\right)-\frac{ \Omega \Delta_\zeta^2(t)}{f_\zeta}\frac{\partial^2}{\partial\zeta^2}(\psi P)-\frac{1}{4} \Omega  f_\zeta\Delta_\zeta^4(t)\frac{\partial^3P}{\partial^2\zeta\partial\psi} \ ,
\label{FP_general}
\end{align}
where we have ignored sub leading terms in the noise, and expanded to cubic order in $t_v$ in order to obtain the last two terms.
Before solving the Fokker-Planck equation let us pay attention to the fact that  $\zeta$ appears only through derivatives of the PDF.  This stems from the fact that $\zeta$ posses a shift symmetry, by which the only allowed interaction contains time derivatives or gradients. As for the stochastic dynamics, this result in the drift being an explicit function of $v_\zeta$. Now, at leading order in $t_v$, we see that this translates into derivatives of $\zeta$. Indeed allowing higher powers of $v_\zeta$ in the drift, translates into higher derivatives of $\zeta$ in the Fokker-Planck equation \cite{Riotto:2011sf,Cohen:2021jbo}. 
Another important feature of Eq.~\eqref{FP_general} is that, as expected, in the absence of the coupling $\Omega$ the equation reduces to uncoupled Fokker-Plank equations whose solution factorises as $P\sim P_{\psi}(\psi,t)P_{\zeta}(\zeta,t)$.

Finally let us stress that in order to obtain Eq. \eqref{FP_general} we only need to assume that $f_\zeta^2\gg H^2$,\ $\Omega^2\ll H^2$, and $\mu^2\ll H^2$. The assumption about $\Omega$ can be relaxed, but  doing that will modify the value of the two point function which will modify the noise function.

\subsubsection*{Including higher $\zeta$ derivatives}

Let us now study the last two terms in \eqref{FP_general} and restrict to $V(\psi)=\mu^2 \psi^2/2$. These contain two derivatives of $\zeta$, and as expected will modify the variance of the distribution for $\zeta$. As we will see they restrict the range of the Gaussian fluctuations for $\psi$. Indeed we can find the solution for the PDF at equilibrium to be
\begin{align}
\log P(\zeta,\psi)\propto -\frac{\psi^2}{2\sigma_\psi^2}-\frac{\left(\zeta+\frac{2\Omega f_\zeta}{3H}\frac{\sigma_\zeta^2}{\sigma_\psi^2}\psi-\frac{\Omega^2}{3H^2}\frac{\sigma_\zeta^2}{\sigma_\psi^2}\psi^2\right)^2}{\sigma_\zeta^2(1+\frac{3\Omega f_\zeta}{t_\psi H}\frac{\sigma_\zeta^2}{2\sigma_\psi^2}\psi)} \ ,
\label{other_interactions_PDF}
\end{align}
where we have neglected other terms that appear in the denominator which are suppressed by additional powers of $\Delta_\zeta^2$. We see from \eqref{other_interactions_PDF} that the variance gets shifted for larger values of $\psi$. Notice, moreover,  that for typical values of $\psi$ this effect is very suppressed since $\Omega\ll f_\zeta$. In any case, at that point the  distribution is not valid and other terms will become dominant. This effect would modify the tail of $\psi$, and as long as we consider small fluctuations around the new trajectory of $\psi$ the description we have given for the distribution in $\zeta$ remains valid. If we  expand  in small $\psi$ we find that there are new terms proportional to $\psi^2\zeta^2$. These terms would actually change the saddle point moving it back to $\psi= 0$ at larger values of $\zeta$. This effect is related to the condition we impose in \eqref{Neglecting_vz_drift}. In the end we  find that  a violation of \eqref{Neglecting_vz_drift}  is related to the fact that  there is no change in the saddle points due to the quadratic term in $\psi$.  In general we found that for late times and larger $\Omega^2/H^2$ the effect of these new terms is suppressed. The reason behind this is that for late times $t\gtrsim t_\psi$ the variance of $\sigma_\psi$ stops growing whereas the variance of $\zeta$ grows until the end of inflation.

\subsubsection*{General $V(\psi)$}

We have focused only on quadratic potentials for $\psi$ but it possible to study a general potential. To start with, let us neglect the linear coupling and  focus only on the leading order interactions.  If we Fourier transform in $p$ and write $P=\exp(-\sigma_\zeta^2/2p^2)\tilde{P}$ we obtain a  Fokker-Planck equation with only  derivatives of $\psi$ and with a corrected drift term
\begin{align}
\frac{\partial\tilde P}{\partial t }=\frac{\partial}{\partial\psi}\left(\left(\frac{V'(\psi)}{3H}+ip \frac{2\Omega^2\Delta_\zeta^2(t)}{H}\psi \right)\tilde P\right)-\frac{D_\psi}{2}\frac{\partial^2 \tilde P}{\partial\psi^2} \ .
\end{align} 
We can rewrite  this equation using the replacement
\begin{align}
\tilde P(\zeta,\psi,t)=e^{-v/2}\sum_{n=0}^{\infty}a_n\Phi_n e^{\Lambda_n t},\qquad v(\zeta,\psi,t)=\frac{8\pi^2}{3H} \left(V(\psi)+3ip\frac{\Omega^2}{H^2}\Delta_\zeta^2(t)\psi^2\right) \ ,
\end{align} by which the Fokker-Planck equation becomes an eigenvalue problem of the form
\begin{align}
\frac{1}{2}\left(-\frac{\partial}{\partial\psi}+\frac{\partial v}{\partial\psi}\right)\left(\frac{\partial}{\partial\psi}+\frac{\partial v}{\partial\psi}\right)\Phi_n(\psi)=\frac{8\pi^2\Lambda_n}{H^3}\Phi_n(\psi) \ ,
\end{align}
which is similar to Eq. \eqref{eq:eigenvalue_FP} with the particular difference that the equilibrium distribution $\exp(-v)$ in this case is time independent. Solving this equation in general lies beyond the scope of this paper, but we can still comment on the case that $\psi$ reaches the equilibrium PDF. Fourier transforming in $p$ we find that the PDF is given by
\begin{align}
P(\zeta,\psi)\sim\exp\left[ -\frac{8\pi^2 V(\psi)}{3H^2}-\frac{1}{2\sigma_\zeta^2}\left(\zeta-\frac{\bar\kappa'}{2\sigma_{0\psi}^2}\psi^2\right)^2\right] \ ,
\end{align}  
which is a generalisation of \eqref{Joint_distribution_equilibrium}, including general potentials. Let us consider the example $V(\psi)=\mu^2 \psi^2 /2+\lambda\psi^4/4$ which allows us to integrate $\psi$ out of the equilibrium distribution. This leads to
\begin{figure}
  \includegraphics[scale=0.5,trim=1.8cm 9cm 1cm 4cm,clip]{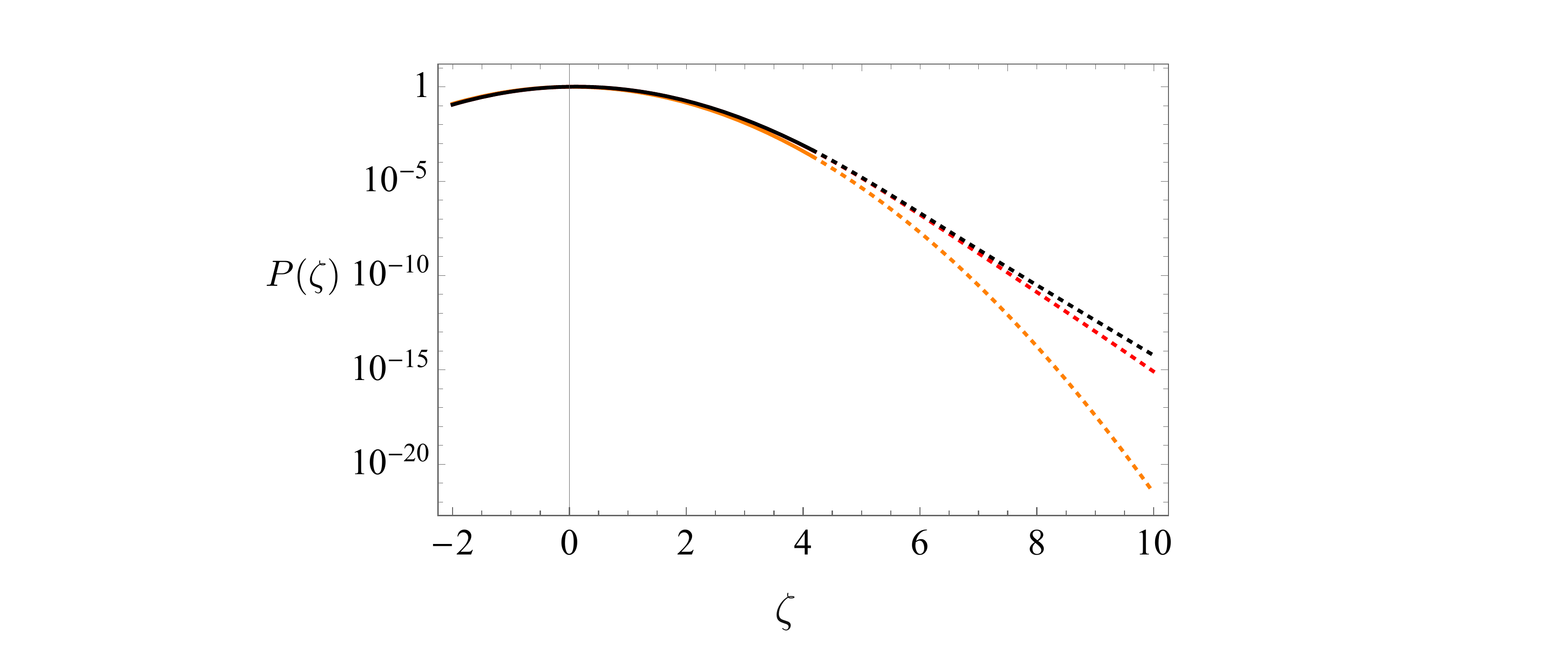}
\caption{We plot the log of the $\zeta$ distribution given by \eqref{cases:eq_lambda} for three different values of $g$. The black line corresponds to $\lambda=0$, the red line to $\lambda=10^{-5}$ and the orange line to $\lambda=10^{-2}$. The dotted lines are the regions where $ 3\zeta> Ht\frac{\mu^2}{\Omega^2}$.}
\label{Fig:tails_lambda}
\end{figure}
\begin{align}
P(\zeta)\propto
\begin{cases}
\exp\left(\frac{1}{4\sigma_\zeta^2}\frac{(\zeta-\sigma_\zeta^2/\bar\kappa')^2}{1+\frac{6\lambda t_\psi\sigma_\psi^2\sigma_\zeta^2}{H\bar\kappa'^2}}-\frac{\zeta^2}{2\sigma_\zeta^2}\right)K_{1/4}\left(\frac{1}{4\sigma_\zeta^2}\frac{(\zeta-\sigma_\zeta^2/\bar\kappa')^2}{1+\frac{6\lambda t_\psi\sigma_\psi^2\sigma_\zeta^2}{H\bar\kappa'^2}^2}\right) \\
\hspace{8.4cm}\mathrm{for}\ \zeta<\frac{\sigma^2_\zeta}{\bar\kappa'}\ ,\\
\\
\exp\left(\frac{1}{4\sigma_\zeta^2}\frac{(\zeta-\sigma_\zeta^2/\bar\kappa')^2}{1+\frac{6\lambda t_\psi\sigma_\psi^2\sigma_\zeta^2}{H\bar\kappa'^2}}-\frac{\zeta^2}{2\sigma_\zeta^2}\right)\left(I_{-1/4}\left(\frac{1}{4\sigma_\zeta^2}\frac{(\zeta-\sigma_\zeta^2/\bar\kappa')^2}{1+\frac{6\lambda t_\psi\sigma_\psi^2\sigma_\zeta^2}{H\bar\kappa'^2}}\right)+I_{1/4}\left(\frac{1}{4\sigma_\zeta^2}\frac{(\zeta-\sigma_\zeta^2/\bar\kappa')^2}{1+\frac{6\lambda t_\psi\sigma_\psi^2\sigma_\zeta^2}{H\bar\kappa'^2}}\right)\right)\\
\hspace{8.4cm}\mathrm{for}\ \zeta>\frac{\sigma^2_\zeta}{\bar\kappa'}\ .
\end{cases}
\label{cases:eq_lambda}
\end{align}
This expression is a bit complicated but we can  see that a quartic self interaction reduces the amplitude of the tail. This has an overall effect of making the distribution Gaussian for larger values of $\zeta$. We plot \eqref{cases:eq_lambda} in   Fig.~\ref{Fig:tails_lambda} where it is noticeable how the non Gaussian effects diminish by increasing $\lambda$.
This can be shown by expanding around the non trivial saddle point. In this case after integrating out $\psi$ we find that
\begin{align}
P(\zeta)\sim\exp\left(-\frac{\zeta^2}{\frac{H\bar\kappa^2}{3t_\psi\lambda\sigma_\zeta\sigma_\psi^2}+2\sigma_\zeta^2}-\frac{\zeta}{\bar\kappa+\frac{6t_\psi\lambda\sigma_\psi^2\sigma_\zeta^2}{H \bar\kappa}}\right) .
\end{align}
Notice that the tail contains a quadratic term, whose variance has a correction which is the inverse of the correction of the coefficient of the tail. This means that increasing $\lambda$ makes the correction of the quadratic term smaller while the correction for the linear term becomes larger. The effect can be understood by estimating the sizes of the non Gaussianites. Indeed we have that for the quartic interaction $f_{\mathrm{NL}}^{(\lambda)}$, is of order
\begin{align}
f_{\mathrm{NL}}^{(\lambda)}\ \zeta\sim \frac{\lambda}{H^2f_\zeta^2}\frac{\sigma_\psi^4}{\sigma_\zeta^2}\times \zeta  .
\end{align} 
If we compare the ratio between $f_{\mathrm{NL}}^{(\lambda)}$ and  $f_{\mathrm{NL}}$ we find that it coincides  with the ratio that controls whether the Gaussian term dominates
\begin{align}
\frac{f_{\mathrm{NL}}^{(\lambda)}}{f_{\mathrm{NL}}}\sim\frac{\lambda}{\Omega^2}\sigma_\psi^2 .
\label{fnl:ratio}
\end{align}
Let us first detail the case in which the ratio is much larger than one.
When this  happens, since perturbation theory breaks down for $f_{\mathrm{NL}}^{(\lambda)}\zeta\sim 1$, the action cannot be trusted anymore and the effect of the cubic coupling between $\zeta$ and $\Omega$ is not seen. Clearly at this point other self interactions have to be taken into account and the PDF at large $\zeta$ might  become dominated by other higher order  terms.
On the other hand when the ratio in \eqref{fnl:ratio} becomes smaller than one the saddle point changes before $f_{\mathrm{NL}}^{(\lambda)}\zeta\sim 1$ and so the tail  becomes  non Gaussian. For any non zero $\lambda $ there it will be a point when  $f_{\mathrm{NL}}^{(\lambda)}\zeta\sim 1$. At that stage the computations of the tail in are not valid. Nevertheless this will produce an exponentially small effect on the whole PDF.

\section{Conclusions}
\label{sec:conclusions}

We have studied the statistics of large but rare fluctuations within the multi-field inflation paradigm using the stochastic inflation formalism. In the simplest class of two-field models, the primordial curvature fluctuation $\zeta$ interacts with an isocurvature field $\psi$ as a result of turns of the background trajectory in the target space of scalar fields. This translates into a derivative coupling proportional to $\Omega$, the rate of turn of the trajectory, appearing at both linear and non-linear level in the evolution of perturbations. We found that the non-linear interactions induced by $\Omega$ imply non-Gaussian deformations affecting the tails of the joint probability distribution of the perturbations. 

By assuming that the evolution of the background is close to de Sitter, we derived the Fokker-Planck equation that is satisfied by the probability density function characterising both fields $\zeta$ and $\psi$. When only the linear evolution of the fields is considered, we find that a non vanishing $\Omega$ enhances the growth of the variance of $\zeta$. This matches with results obtained previously using perturbation theory. A particular case of this scenario is when the entropy mass is exactly zero (the ultralight limit), studied recently in Ref.~\citep{Achucarro:2016fby}. We showed that initially all spectator fields behave as ultralight fields after horizon crossing but after some time, that depends on their mass, the fields decay. If the entropy mass is zero, we recover the exact ultralight case. 

On the other hand, when non linearities are taken into account, we find that after integrating out $\psi$ the tail of the PDF of $\zeta$ becomes non-Gaussian for values of $\zeta \gtrsim H^2/\Omega^2$. This can be understood as the Stokes phenomenon, whereby a Gaussian saddle point leads to non-Gaussian saddle points for large values of a parameter. Crucially the coupling makes the exponential tail to be   larger than the Gaussian tail. Such a result has important consequences. For instance the abundance of PBHs formed during inflation depends strongly on the tail of the PDF. In this way a non-Gaussian tail implies that the abundance of PBHs can be substantially enhanced in models with derivative couplings. Another consequence would be a possible  modification on the clustering of galaxies which depend on rare large fluctuations, whose probability would now be enhanced.  

There are several paths along which our work can be expanded. For instance, here we considered a simple EFT of multi field inflation in which the number of interaction terms with derivatives couplings is limited. There are other known examples with a larger number of interaction terms which might become relevant for large values of $\zeta$. In those cases one might need to resum the implied corrections to obtain accurate expressions for the tail. Another interesting topic would be to understand how corrections to the Fokker-Planck equation are related to resummation of loops. This has been well understood for light spectator fields on pure de Sitter \cite{Starobinsky:1994bd,Baumgart:2019clc,Gorbenko:2019rza,Green:2020jor}, and a similar result should follow from the Fokker-Planck equation arising from our two field model. 
Also, it would be interesting to compare our results to other derivations of the Fokker-Planck equations within the multi-field paradigm~\cite{Salopek:1990re,Assadullahi:2016gkk,Vennin:2016wnk,Pinol:2020cdp} and its relation with the Hamilton-Jacobi formalism \cite{Salopek:1990jq,Achucarro:2019pux}. On these examples the derivation of the Fokker-Planck equation was done directly from the background equations, whereas in this article we obtained the stochastic dynamics directly from perturbations.

Our results suggest that the abundance of PBHs in multifield models can be much larger than that obtained when the Gaussian approximation is used to study the production of PBHs through the enhancement of the power spectrum. It would be interesting to apply our results to models such us those of Refs.~\cite{Palma:2020ejf,Fumagalli:2020adf} to reassess the production of PBHs. To do this one would need to go beyond the assumption that $\Omega\ll H$, implying that some of the terms that in our analysis were suppressed would now become dominant. 
Finally we see that stochastic inflation might allow one to go beyond perturbation theory. In that sense it will be interesting to understand our results in the light of recent works such as~\cite{Celoria:2021vjw, Cohen:2021fzf} (see \cite{Fumagalli:2020nvq, Ballesteros:2021fsp} for a discussion on the importance of this for PBHs). For instance  it has been suggested that when there is a non perturbative tail  there is an exponential enhancement of the large $N$ point correlation function \cite{Panagopoulos:2020sxp}. Whether those result apply to the case we study here, we leave  for future work.

\section*{Acknowledgements}

We are grateful to Guillermo Ballesteros, Lucas Pinol and Spiros Sypsas for useful discussions and comments on this work. G.A.P is supported by the Fondecyt Regular Project No.~1210876 (ANID). The work of  SC has been funded by a Contrato de Atracci\'on de Talento (Modalidad 1) de la Comunidad de Madrid (Spain), number 2017-T1/TIC-5520 and the IFT Centro de Excelencia Severo Ochoa Grant SEV-2016-0597. ACD acknowledges partial support from STFC consolidated grant ST/T000694/1. AA's work is partially supported by the Netherlands Organization for Scientific Research (N.W.O), by the Basque Government (IT-979-16) and by the Spanish Ministry MINECO (FPA2015-64041-C2-1P).

\appendix

\section{Langevin equations}
In this appendix we give details on how to obtain the Langevin equation (\ref{eq:pre-LangevinSFI}). Starting from the action
\begin{align}
S=\int d^4 x dt a^3\epsilon\left[\dot\zeta^2-\frac{1}{a^2}(\nabla\zeta)^2\right] \ ,
\end{align}
the equations of motion are
\begin{align}
\ddot \zeta+3H\dot\zeta+\frac{k^2}{a^2}\zeta=0 \ .
\label{eq:eomzeta}
\end{align}
We can split the solution to this equation into long- and short-wavelengths
\begin{align}
\zeta(x)&=\zeta_l(x)+\zeta_s\nonumber\\
		&=\zeta_l+\int\frac{d^3k}{(2\pi)^3}\theta(k-\epsilon a(t)H)\left[a_k\zeta_k(t)e^{-i \vec{k}\cdot\vec{x} }+a_k^{\dagger}\zeta^*_k e^{i \vec{k}\cdot\vec{x} }\right] \ ,
\end{align}
with $\epsilon\ll 1$ and where $\zeta_k$ is the solution of \eqref{eq:eomzeta}
\begin{align}
\zeta_k=z\sqrt{\frac{\pi}{4}}H(-\tau)^{3/2}H_{3/2}^{(1)}(-k \tau) \ ,
\end{align}
where $a,a^{\dagger}$ are the corresponding creation and annihilation operators and $\tau$ is conformal time. Notice that picking a different window function might change the result of the correlation function (see \cite{Winitzki:1999ve} for a discussion) although in this work  we are only interested in equal-time correlation functions. Clearly the definition of the long wavelength modes implies that we neglect the gradient term in the equation of motion. Moreover, since horizon  crossing happens at $k/aH\simeq 1$, it also implies that the leading piece from the equation of motion is $3H\dot\zeta_l$. Due to this we  can write the equation of motion as
\begin{align}
\ddot\zeta_l+ 3H\dot \zeta_l=\eta_\zeta(t,x) \ ,
\end{align}
where $\eta_\zeta(t,x)$ is the contribution from the short wavelength perturbations.

\subsection{Computing the noise terms}
In order to compute the noise term let us consider a light scalar field $\phi$ in de Sitter space of mass $m^2\ll H^2$ . Going back to $\zeta$ can be done by rescaling our results. Following \cite{Starobinsky:1986fx}, let us split the field into long and short wavelengths by writing it as
\begin{align}
\phi(x)&=\phi_l(x)+\phi_s\nonumber\\
		&\equiv\bar{\phi}+\int\frac{d^3k}{(2\pi)^3}\theta(k-\epsilon a(t)H)\left[a_k\phi_k(t)e^{-i \vec{k}\cdot\vec{x} }+a_k^{\dagger}e^{i \vec{k}\cdot\vec{x} }\right] \ , \label{eq:smearing}
\end{align}
where we have introduced the time dependent cut-off $k_c=\epsilon a(t) H$, with $\epsilon$ a small parameter, and where the mode function $\phi_k$ is given by
\begin{equation}
\phi_k=\sqrt{\frac{\pi}{4}}H(-\tau)^{3/2}H_{\nu}^{(1)}(-k \tau),
\end{equation} where $\tau$ is conformal time and $\nu=\sqrt{9/4-m^2/H^2}$ . The noise term comes from averaging over the time derivative of the second term. Indeed defining
\begin{equation}
f\equiv \epsilon a(t)H^2\int\frac{d^3k}{(2\pi)^3}\delta(k-\epsilon a(t)H)\left[a_k\phi_k(t)e^{-i \vec{k}\cdot\vec{x} }+a_k^{\dagger}e^{i \vec{k}\cdot\vec{x} }\right] \ ,
\end{equation}
we then have that
\begin{align}
\langle f(x,t)f(x',t')\rangle &=\epsilon^2 H^{4}\int \frac{d^3k}{(2\pi)^3}\phi_k\phi^*_k e^{-i \vec{k}\cdot (\vec{x}-\vec{x}')}\delta(k-\epsilon a(t)H) a(t)\delta(k-\epsilon a(t')H)a(t')\nonumber\\
&=\epsilon^2 H^4 \int \frac{k^2 d k}{2\pi^2}\phi_k\phi_k^*\frac{\sin(k \vert\vec{x}-\vec{x}'\vert)}{k \vert\vec{x}-\vec{x}'\vert}\delta(k-\epsilon a(t)H) a(t)\delta(k-\epsilon a(t')H)a(t').\nonumber\\
\label{eq:SW_correlation}
\end{align}
Now for a massless field we have that
\begin{equation}
\phi_k=\frac{H}{\sqrt{2k}}\left(\tau-\frac{i}{k}\right)e^{-i k\tau},
\end{equation}
so instead we obtain
\begin{align}
\langle f(x,t)f(x',t')\rangle &= \frac{\epsilon^2 H^4 }{4\pi^2}\frac{\sin (\epsilon a(t)H\vert\vec{x}-\vec{x}'\vert)}{\epsilon a(t)H\vert\vec{x}-\vec{x}'\vert}\frac{a(t)^2}{\epsilon a(t) H} \frac{\delta(t-t')}{\epsilon H^2 a(t)}\nonumber\\
&=\frac{H^3}{4\pi^2}\frac{\sin (\epsilon a(t)H\vert\vec{x}-\vec{x}'\vert)}{\epsilon a(t)H\vert\vec{x}-\vec{x}'\vert}\delta(t-t'),
\end{align}
which implies that at coincident points we have
\begin{equation}
\langle f(x,t)f(x,t')\rangle =\frac{H^3}{4\pi^2}\delta(t-t').
\end{equation}
In general we should expand $\phi_k\phi_k^*$ in small $\epsilon$ after replacing the cut-off in the mode functions. At leading order in $\epsilon$ we find that
 \begin{equation}
\phi_k\phi_k^*=\frac{\Gamma(\nu)^2 4^{\nu-1}}{\pi }\frac{\epsilon^{-2\nu}}{H a^3}.
\end{equation}
Replacing into \eqref{eq:SW_correlation} we get
\begin{align}
\langle f(x,t)f(x',t')\rangle &= \frac{\epsilon^2 H^4 }{2\pi^2}\frac{\sin (\epsilon a(t)H\vert\vec{x}-\vec{x}'\vert)}{\epsilon a(t)H\vert\vec{x}-\vec{x}'\vert}\epsilon^2 a(t)^2H^2\frac{\Gamma(\nu)^2 4^{\nu-1}}{\pi }\frac{\epsilon^{-2\nu}}{H a^3} \frac{a(t)^2\delta(t-t')}{\epsilon H^2 a(t)}\nonumber\\
&=\frac{H^3}{4\pi^2}\frac{\sin (\epsilon a(t)H\vert\vec{x}-\vec{x}'\vert)}{\epsilon a(t)H\vert\vec{x}-\vec{x}'\vert}\frac{\Gamma(\nu)^2 4^{\nu-1/2}}{\pi}\epsilon^{3-2\nu}\delta (t-t'),
\end{align}
Expanding in powers of $m/H$ we get,
\bea
\langle f(x,t)f(x',t')\rangle &=& \frac{H^3}{4\pi^2}\frac{\sin (\epsilon a(t)H\vert\vec{x}-\vec{x}'\vert)}{\epsilon a(t)H\vert\vec{x}-\vec{x}'\vert} \nonumber \\ &&
\!\!\!\!\!\!
\times \left(1+2\left(-2+\gamma_E+\log (2\epsilon)\right)\frac{m^2}{H^2}+\mathcal{O}\left(m^2/H^2\right)\right)\delta (t-t'). \quad \label{eq:noiseterm_general}
\eea

\section{Secular growth of $\sigma_{\zeta}^2$}

In this appendix we will relate the variances of the stochastic fields to the correlation functions in real space. First let us recall that the field perturbation in inflation can be written approximately as
\begin{align}
    \delta\phi_k=\sqrt{\frac{\pi}{4}} H (-\tau)^{3/2}H_{3/2}^{(1)}(-k\tau),
\end{align} 
where $\tau$ is conformal time and $a(\tau)=-1/(H\tau)$.
From this it is possible to write the two point function in momentum space $\langle \delta\phi_k^2\rangle=\frac{H^2}{2 k^3}(1+k^2\tau^2)$. Since we would like to compare this to the result obtained through the Langevin equation let us Fourier transform the two point function. We have that
\begin{align}
    \langle\delta\phi(\vec{x},t)\delta\phi(\vec{y},t)\rangle&=\int\frac{d^3k}{(2\pi)^3}e^{i \vec {k}\cdot\vert\vec{x}-\vec{y}\vert}\langle\delta\phi_k^2\rangle\nonumber\\
    &=\frac{H^2}{4\pi^2}\int k^2 dk e^{i k\vert\vec{x}-\vec{y}\vert}\frac{1}{ k^3}(1+k^2\tau^2)\nonumber\\
    &=\frac{1}{(2\pi)^2}\frac{1}{\vert\vec{x}-\vec{y}\vert^2 a(t)^2}-\frac{H}{(2\pi)^2}\log\left(\frac{\vert\vec{x}-\vec{y}\vert}{L}\right),
    \label{eq:2ptfunction}
\end{align}
where the infrared cut-off $L\equiv a_0 H_0$ is the largest scale during inflation. This is related to the end of inflation as we can write $(a_0 H_0)^{-1}=-\tau_0>0$ where $\tau_0<0$ is the time when inflation ends. Notice  also that during eternal inflation $\tau_0\to 0$ \cite{Creminelli:2008es}.

Eq.~\eqref{eq:2ptfunction} can be understood as follows,  the first piece corresponds to the flat space two point function in physical coordinates. As we move deeper into the bulk this expression dominates. This is expected as we have picked a vacuum that reproduces the Minkowski vacuum. This term dilutes as we approach the horizon. The second term does not depend explicitly on time and it corresponds to   the two point function on a scale invariant theory. This is due to the symmetries of inflation at horizon crossing.

In order to compare to the stochastic result let us compute the correlation functions at coincident points. This corresponds to the variance of $\delta\phi$. Since the expression diverges let us introduce a time dependent cut-off such that 
\begin{align}
    \vert\vec{x}-\vec{y}\vert e^{Ht}\gg \Lambda^{-1}.
\end{align}
For shorter distances  we evaluate the two point at the cut-off, which schematically implies that
\begin{align}
    \langle\delta\phi(t)^2\rangle&=a \Lambda^2+b\log\Lambda+\frac{H^3 t}{(2\pi)^2}=\frac{H^3 t}{(2\pi)^2}+\mathrm{const},
\end{align}
which is the result we have obtained by solving the Fokker-Planck equation. Notice that the time dependence comes from the fact that at each time more modes are included in the region below the cut off. 
A similar computation shows that at coincident points the $\dot\phi$ correlation function goes as
\begin{align}
    \langle \delta\dot\phi(\vec x,t)\delta\dot\phi(\vec x,t')\rangle_{\rm{vac}}&=\frac{1}{2\pi^2}\int_0^\infty d^3k \vert\delta\dot\phi_k(\tau)\delta\dot\phi_k(\tau')\vert\nonumber\\
    &=\frac{6H^4}{4\pi^2}\frac{\tau^2\tau'^2}{(\tau-\tau')^4}\nonumber\\
    &=\frac{3H^4}{32\pi^2}\left(\sinh(H\vert t-t'\vert\right)^{-4},
    \end{align}
where we have pointed out that these are vacuum fluctuations, to distinguish them from the statistically averaged two point functions. 
To compute this, let us note that  after smearing  the field  the noise function  is directly proportional to the smeared speed, as can be seen after taking the time derivative from  \eqref{eq:smearing}. Then, we have that the correlation function is given by
\begin{align}
     \langle \delta\dot\phi(\vec x,t)\delta\dot\phi(\vec x,t')\rangle_{\mathrm{av}}&=\frac{H^4}{8\pi^2}\delta(t-t').
\end{align}
Notice that, while at separate time both decay to zero, at equal times the correlation function diverges while the statistical average is finite. This is an effect of smearing out over a region where there are statistical fluctuations, which in the end translates into the correlation function for the speed being finite at equal time.

\section{General solution for linear coefficients}
\label{sec:General_solution}
In this appendix we will  follow~\cite{van1992stochastic}  to derive general solutions of the Fokker-Planck equation. Let us start by considering the following equation,
\begin{equation}
\frac{\partial P}{\partial t}=-A_{ij} \frac{\partial}{\partial\phi_i}(\phi^j P)+\frac{1}{2}D_{ij}\frac{\partial^2}{\partial\phi_i\phi_j}P \ ,
\end{equation}	where both  $A_{ij}$ and $D_{ij}$ are $n\times n$ constant matrices and in addition $D_{ij}$ is symmetric and semipositive definite. Subject to initial conditions
\begin{equation}
P(\phi,0)=\prod_i \delta(\phi^i-\phi^i_0).
\end{equation}
The solution of this equation is Gaussian which we will show. First if we  multiply the equation for $\phi^i$ and integrate over $\phi$ we find after integration by parts,
\begin{equation}
\partial_t\langle \phi^i\rangle= A_{kj}\langle\phi^j\rangle \ ,
\end{equation}
whose solution is given by \begin{equation}
\langle \phi^i\rangle=e^{t A}y_0 \ ,
\end{equation}
in matrix notation.  Now if we insert $\phi^i\phi^j$ into the Fokker-Planck equation we find,
\begin{equation}
\partial_t\langle\phi^i\phi^j\rangle=A^i_{k}\langle\phi^k\phi^j\rangle+A^j_{k}\langle\phi^k\phi^i\rangle+D^{ij} \ .
\end{equation}
It is more convenient to use covariance matrix $C_{ij}=\langle\phi_i\phi_j\rangle-\langle\phi_i\rangle\langle\phi_j\rangle$. In which case the above equation reduces to,
\begin{equation}
\partial_t C= A C+ C A^{t}+D \ ,
\end{equation}
in matrix notation. Now if we write the covariance matrix as $C=e^{t A}\bar{C}e^{t A^t}$ then we get
\begin{equation}
\partial_t \bar{C}=e^{-t A}D e^{-t A^t} \ ,
\end{equation}
where assuming that $\bar{C}(0)=0$ has a solution given by,
\begin{equation}
C(t)=\int_0^t e^{(t-t')A}D e^{(t-t')A^t}dt' \ .
\end{equation}
Even though this expression looks abstract it can be easily computed . To conclude given that we assume that the distribution was Gaussian then it is fully determined by the covariance matrix,  hence we have,
\begin{equation}
P(\phi,t)=\frac{1}{(2\pi)^{n/2}}\frac{1}{\sqrt{\det C}}\exp\left(\frac{1}{2}(\phi-\langle\phi\rangle)^t\rangle  C^{-1}(\phi-\langle\phi\rangle )\right) \ .
\end{equation}
Let us check now that the expression for $C$ agrees with what we found before. We first have that,
\begin{align}
A=\begin{pmatrix}
0 & 1 \\
0 & -3H
\end{pmatrix},\qquad D=\begin{pmatrix}
0 &  0\\
0 & D_\zeta
\end{pmatrix}.
\end{align}
The eigenvalues of $A$ are $0$ and $3H$, and we can compute the exponential of $A$
\begin{align}
\exp(-t A)=\left(
\begin{array}{cc}
 1 & \frac{1-e^{-3 H t}}{3 H} \\
 0 & e^{-3 H t}  \\
\end{array}
\right),
\end{align}
We then have that,
\begin{align}
 e^{(t-t')A}D e^{(t-t')A^t}=\left(
\begin{array}{cc}
 \frac{D \left(e^{3 H \left(t'-t\right)}-1\right)^2}{9 H^2} & \frac{D e^{6 H
   \left(t'-t\right)} \left(e^{3 H \left(t-t'\right)}-1\right)}{3 H} \\
 \frac{D e^{6 H \left(t'-t\right)} \left(e^{3 H \left(t-t'\right)}-1\right)}{3 H} & D
   e^{6 H \left(t'-t\right)} \\
\end{array}
\right).
\end{align}
Performing the integral we find ,
\begin{align}
C(t)=\left(
\begin{array}{cc}
 -\frac{D \left(-6 H t+e^{-6 H t}-4 e^{-3 H t}+3\right)}{54 H^3} & \frac{D e^{-6 H t}
   \left(e^{3 H t}-1\right)^2}{18 H^2} \\
 \frac{D e^{-6 H t} \left(e^{3 H t}-1\right)^2}{18 H^2} & -\frac{D \left(e^{-6 H
   t}-1\right)}{6 H} \\
\end{array}
\right),
\end{align}
which coincides with the expression we found before. Taking the limit $t \gg 1/H$ we have
\begin{equation}
C(t) \to \left(
\begin{array}{cc}
 \frac{D t}{9 H^2} & \frac{D}{18 H^2} \\
 \frac{D}{18 H^2} & \frac{D}{6 H} \\
\end{array}
\right),
\end{equation}
so we find that 
\begin{equation}
P=\frac{1}{2\pi}\sqrt{\frac{54 H^3}{D_\zeta^2 t}}\exp\left(-\frac{9 H^2}{2D_\zeta t}\zeta^2+\frac{3H}{D_\zeta t}v_\zeta \zeta-\frac{3H}{D_\zeta} v_\zeta^2\right) \ .
\end{equation}
\section{Ultralight field}
\label{UL_section}
In this appendix we will show how to modify the noise term in the Fokker-Planck equation to take into account the superhorizon time dependence of the curvature power spectrum. This can be done in general but in this case we will focus on the case of an ultralight field \cite{Achucarro:2016fby}. Let us start from the Langevin Eqs. \eqref{eq:linearisedLangevin}, specialised to the case of $\mu^2=0$.
 For convenience we will define a new variable $\tilde \zeta$ such that the Langevin equations are now,
\begin{align}
\dot{\tilde \zeta}&=v_\zeta,\nonumber\\
\dot v_\zeta&=-3H v_\zeta+\eta_{\tilde\zeta}(t),\nonumber\\
\dot  \psi&=v_\psi,\nonumber\\
\dot \psi&=-3Hv_\psi+2\Omega f_\zeta v_\zeta+\eta_\psi(t).
\end{align}
Notice that in this case $\tilde\zeta$ grows outside the horizon due to the interactions with $\psi$. This implies that the noise term $\eta_{\tilde\zeta}(t)$, which is computed by coarse graining $\tilde \zeta$, should also grow on superhorizon scales. To do so, let us first  recall that for an ultralight field, the two point function for the curvature mode is given by~\cite{Achucarro:2016fby}. ,
\begin{equation}
\langle \zeta^2\rangle=\frac{1}{2\epsilon a^2}\frac{1}{2k^3\tau^2}\left(1+\lambda^2\left[A_1-A_2\log(-k\tau)+\log^2(-k\tau)\right]\right) \ ,
\end{equation}
 where $\lambda\equiv \frac{2\Omega}{H}$, and  $A_1$, $A_2$ are given by,
\begin{align}
A_1&=-\frac{\pi^2}{6}+(3-\ln 2)(1-\ln 2)-\gamma_E(4-\gamma -2\log 2)\simeq -2.11 \ ,\nonumber\\
A_2&=4-2\gamma -2\ln 2\simeq 1.46 \ .
\end{align}
Notice that it diverges in the limit $k \tau\to 0$. However, since inflation will last for a finite amount of time, then there is an natural cut-off for the power spectrum.  A more systematic  way to deal with this IR behaviour is to regularise the growing logs by introducing boundary counterterms as in ~\cite{Cespedes:2020xqq}. Doing so results in the following regularised expression,
\begin{equation}
\langle \zeta^2\rangle=\frac{1}{2\epsilon a^2}\frac{1}{2k^3\tau^2}\left(1+\lambda^2\left[A_1-A_2\log(k/\mu_{IR})+\log^2(k/\mu_{IR})\right]\right),
\end{equation}
where $\mu_{IR}$ is an infrared cut-off. If we set $\mu_{IR}=\epsilon H$ the diffusion coefficient for $\zeta$ changes to,
\begin{align}
D_\zeta\to D_\zeta\left(1+\lambda^2\left[A_1-A_2\log(a)+\log^2(a )\right]\right).
\end{align}
To compute the covariance matrix we can use the results from Appendix \ref{sec:General_solution}, which are compatible with a time dependent difussion matrix. At leading order in $\lambda^2 \Delta N^2 $, and for $t\gg 1/H$, we have that,

\begin{align}
C=\left(
\begin{array}{cccc}
 \frac{Dt(3+ H^2 t^2\lambda^2)}{27 H^2} & -\frac{\sqrt{2\epsilon}Dt(3+H^2 t^2\lambda^2)}{81 H^2} & \frac{D}{18 H^2} & 0 \\
 -\frac{\sqrt{2\epsilon}Dt(3+H^2 t^2\lambda^2)}{81 H^2}  & \frac{H^3 t}{4\pi^2} & 0 & \frac{H^3}{8\pi^2} \\
 \frac{D}{18 H^2} & 0 & \frac{D}{6 H} & 0 \\
 0 & \frac{H^3}{8\pi^2} & 0 & \frac{3 H^4}{8 \pi ^2} \\
\end{array}
\right)\ ,
\end{align}
where we have kept terms at order $\sqrt{\epsilon}\lambda \Delta N^2$. Notice that this steady state is reached within a couple of efolds. If we  drop the slow roll terms, the PDF is given by

\begin{equation}
P=\frac{6\sqrt{3}}{\sqrt{D_\zeta^2H^6 t^4\lambda}}\exp\left(-\frac{27 H^2}{2 D_\zeta t^3\lambda^2}\tilde\zeta^2+\frac{9}{D_\zeta H t^3\lambda^2}\tilde \zeta v_\zeta-\frac{3H}{D}v_\zeta^2-\frac{2\pi^2}{H^3 t}\psi^2+\frac{4\pi^2}{3H^4 t} v_\psi\psi -\frac{4\pi^2}{3H^4}v_\psi^2 \right) \ ,
\end{equation}
It is possible to compute the correlation function directly from the distribution by integrating over all the fields
\begin{align}
\langle\phi^a\phi^b\rangle=\int \prod_i D\phi^i \phi^a\phi^b P \ ,
\end{align}
doing so we get,
\begin{align}
\langle\tilde\zeta^2\rangle=\frac{D_\zeta t^3\lambda^2 }{27},\qquad  \langle\psi^2\rangle=\frac{H^3 t}{4\pi^2} \ .\label{eq:variance_ULF}
\end{align}
These are real space correlation functions, to compare them with the  power spectrum we have to use the relation,
\begin{equation}
\langle\phi^2\rangle=\int d\log k \Delta_\phi^2.
\end{equation} 
Taking derivatives on both sides implies that,
\begin{equation}
\frac{d}{d\log k_*}(\langle\phi^2\rangle)=\Delta_\phi^2,
\end{equation}
where $k_*$ is the horizon crossing wavenumber.
Since  for modes that have crossed the horizon, $t$ can be written as $t_*=\frac{1}{H}\log(k_*/H)$,  we have that
\begin{equation}
\Delta_{\tilde\zeta}^2=\frac{D_\zeta\Delta N^2\lambda^2 }{9 H }=\frac{H^2}{4\pi^2}\alpha^2\Delta N^2,\qquad \Delta_{\psi}^2=\frac{H^2}{4\pi^2} \ ,
\end{equation}
 where we have used  that $\frac{1}{H}\log(k_*/H)=\Delta N$, is the number of efolds until the end of inflation and 
where $D_\zeta= 9 H^5/(8\epsilon\pi^2)$ and $\lambda=\sqrt{2\epsilon \alpha}/H$. This result 
 coincide with the  power spectrum computed in~\cite{Achucarro:2016fby}.
 Finally let us notice that the faster growth in the  variance avoids inflation becoming eternal. This can be seen by the following argument. In general  inflation becomes eternal  if during an interval $t\sim H{-1}$ the quantum fluctuations $\langle\delta\phi^2\rangle^{1/2}$ is larger than the classical change of the field $\Delta\phi=\dot\phi/H$.  From \eqref{eq:variance_ULF} we have that this is avoid if
 \begin{align} \frac{\dot\phi}{H} \sqrt{\frac{D_\zeta \lambda^2}{27 H^3}}=
   \frac{H\lambda}{\sqrt{12}\pi}\leq \frac{\dot\phi}{H} \ .
    \label{eq:ULFineq}
 \end{align}
 where we have used that  at horizon crossing $\zeta=-\frac{H}{\dot\phi}\delta\phi$. The last inequality implies that the condition for eternal inflation is more strict than in single fields inflation (which is that $\sqrt{\Delta_\zeta^2}\leq 1$). Indeed \eqref{eq:ULFineq} can be written as,
 \begin{align}
     \frac{\sqrt{\Delta^2_\zeta}}{\sqrt{3}\Delta{N}}\ll \sqrt{\Delta_\zeta^2}\leq 1 \ .
 \end{align}
Notice that this assumes that the power spectrum didn't vary significantly during the whole inflation. If there is a momentarily increase of the power spectrum, such that $\lambda\gg 1$ then the last inequality will not hold. 
\section{Computing further corrections to the PDF.}

In this appendix we will include higher order corrections to the solution of the Fokker-Planck equation in \eqref{Joint_distribution_Complete}.
First, let us solve the following,
\begin{align}
    \frac{dP}{dt}&=\frac{\partial}{\partial\psi}\left(t_\psi^{-1}\psi P+\frac{D_\psi}{2}\frac{\partial P}{\partial\psi}\right)+\frac{H\Delta_\zeta^2}{2}\frac{\partial^2P}{\partial \zeta^2}\nonumber\\
    &+H^2\Delta_\zeta^2\frac{\partial^2}{\partial_\zeta\partial\psi}\left(\left(\frac{2\Omega^2}{H^2}\psi+\frac{2 f\Omega}{3H}\right)P\right)+\frac{6\Omega^2}{f_\zeta^2 H}\frac{\partial}{\partial\zeta}(\psi^2P) \ ,\label{FP_appA}
\end{align}
Our task will be to add the last term to the PDF \eqref{Joint_distribution_Complete}. To simplify we will look for late time solutions such as $\psi$ has reached its equilibrium distribution. We can eliminate some of the terms in the equation by Fourier transforming $\zeta$ to $p$, and look for  solutions of the form,
\begin{align}
    P(p,\psi,t)\sim\exp(-\sigma_\zeta^2 p^2/2)F(p,\psi)
\end{align}
After replacing into the Fokker-Planck equation \eqref{FP_appA} we obtain,
\begin{align}
    0=&F''(k)+\frac{f_\zeta^2 H}{3D_\psi t_\psi}(6\psi+4i f H t_\psi\Delta_\zeta^2\Omega p +4 i t\psi \Delta_\zeta^2\Omega^2\psi p )F'(\psi) \nonumber\\
    &+\frac{1}{3D_\psi t_\psi}(36 i p t_\psi \psi^2\Omega^2+2f^2 H(3+2i p t_\psi\Delta_\zeta^2\Omega^2))F(\psi) .
    \label{eq:F_psi}
\end{align}
Ignoring the linear mixing term the solution is given by,
\begin{align}
F(p,\psi)&\propto\exp\left(-\frac{\psi^2}{2\sigma_\psi^2}\left(1+\frac{2i p t_\psi\Omega^2\Delta_\zeta^2}{3}-\frac{i}{2}\sqrt{f(p)}\right)\right)\times H_n\left(\sqrt{\frac{\psi^2}{2\sigma_\psi^2}}\left(-\frac{ f(p)}{4}\right)^{1/4}\right).\\
n&=\frac{1}{2}-\frac{-3i+2pt_\psi\Delta_\zeta^2\Omega^2}{\sqrt{f(p)}},\\
f(p,q)&=108 \frac{t_\psi \Omega^2\sigma_\psi^2}{f_\zeta^2 H}p+(3i-2p t_\psi\Delta_\zeta^2\Omega^2)^2 ,
\end{align}
where we have discarded the second solution since it grows for large $\sqrt{p}\psi$.
In order to obtain a simplified expression let us notice that a typical fluctuation of $p\sim 1/\sqrt{\sigma_\zeta^2}$. Using this we can deduce that the at leading order  $f(p)$ is constant, as the ratio between the two leading order terms is given by,
\begin{align}
    108 t_\psi\frac{\Omega^2\sigma_\psi^2}{f_\zeta H}\frac{1}{9}=\frac{12t_\psi^2 \Omega^2\Delta_\zeta^2}{\sqrt{\sigma_\zeta^2}}\ll 1.
\end{align}
This inequality still holds for larger values of the $\zeta$. If we write $f_(p)$ at leading order we find that the Hermite function reduces to one and we recover the usual distribution \eqref{Joint_distribution}.
When adding the linear mixing term the distribution is more complicated but still depends on $f(p)$. If we ignore this term, we find that at leading order the PDF contains further corrections at order $p^2$. The effect of those add up to the quadratic terms that appeared in the drift for $v_\zeta$. In the end this will modify the value of the tail for very  large values of $\zeta$, acting as exponetentially suppressed corrections as expected.

\bibliographystyle{utphys}
\bibliography{bibliography}

\providecommand{\href}[2]{#2}\begingroup\raggedright\begin{thebibliography}{10}

\bibitem{Planck:2019kim}
{\bf Planck} Collaboration, Y.~Akrami {\em et al.}, ``{Planck 2018 results. IX.
  Constraints on primordial non-Gaussianity},''
  \href{http://dx.doi.org/10.1051/0004-6361/201935891}{{\em Astron. Astrophys.}
  {\bf 641} (2020)  A9}, \href{http://arxiv.org/abs/1905.05697}{{\tt
  arXiv:1905.05697 [astro-ph.CO]}}.

\bibitem{Maldacena:2002vr}
J.~M. Maldacena, ``{Non-Gaussian features of primordial fluctuations in single
  field inflationary models},''
  \href{http://dx.doi.org/10.1088/1126-6708/2003/05/013}{{\em JHEP} {\bf 05}
  (2003)  013}, \href{http://arxiv.org/abs/astro-ph/0210603}{{\tt
  arXiv:astro-ph/0210603}}.

\bibitem{Cheung:2007st}
C.~Cheung, P.~Creminelli, A.~L. Fitzpatrick, J.~Kaplan, and L.~Senatore, ``{The
  Effective Field Theory of Inflation},''
  \href{http://dx.doi.org/10.1088/1126-6708/2008/03/014}{{\em JHEP} {\bf 03}
  (2008)  014}, \href{http://arxiv.org/abs/0709.0293}{{\tt arXiv:0709.0293
  [hep-th]}}.

\bibitem{Chen:2009zp}
X.~Chen and Y.~Wang, ``{Quasi-Single Field Inflation and Non-Gaussianities},''
  \href{http://dx.doi.org/10.1088/1475-7516/2010/04/027}{{\em JCAP} {\bf 04}
  (2010)  027}, \href{http://arxiv.org/abs/0911.3380}{{\tt arXiv:0911.3380
  [hep-th]}}.

\bibitem{Chen:2009we}
X.~Chen and Y.~Wang, ``{Large non-Gaussianities with Intermediate Shapes from
  Quasi-Single Field Inflation},''
  \href{http://dx.doi.org/10.1103/PhysRevD.81.063511}{{\em Phys. Rev. D} {\bf
  81} (2010)  063511}, \href{http://arxiv.org/abs/0909.0496}{{\tt
  arXiv:0909.0496 [astro-ph.CO]}}.

\bibitem{Achucarro:2010da}
A.~Achucarro, J.-O. Gong, S.~Hardeman, G.~A. Palma, and S.~P. Patil,
  ``{Features of heavy physics in the CMB power spectrum},''
  \href{http://dx.doi.org/10.1088/1475-7516/2011/01/030}{{\em JCAP} {\bf 01}
  (2011)  030}, \href{http://arxiv.org/abs/1010.3693}{{\tt arXiv:1010.3693
  [hep-ph]}}.

\bibitem{Achucarro:2012yr}
A.~Achucarro, V.~Atal, S.~Cespedes, J.-O. Gong, G.~A. Palma, and S.~P. Patil,
  ``{Heavy fields, reduced speeds of sound and decoupling during inflation},''
  \href{http://dx.doi.org/10.1103/PhysRevD.86.121301}{{\em Phys. Rev. D} {\bf
  86} (2012)  121301}, \href{http://arxiv.org/abs/1205.0710}{{\tt
  arXiv:1205.0710 [hep-th]}}.

\bibitem{Arkani-Hamed:2015bza}
N.~Arkani-Hamed and J.~Maldacena, ``{Cosmological Collider Physics},''
  \href{http://arxiv.org/abs/1503.08043}{{\tt arXiv:1503.08043 [hep-th]}}.

\bibitem{Lee:2016vti}
H.~Lee, D.~Baumann, and G.~L. Pimentel, ``{Non-Gaussianity as a Particle
  Detector},'' \href{http://dx.doi.org/10.1007/JHEP12(2016)040}{{\em JHEP} {\bf
  12} (2016)  040}, \href{http://arxiv.org/abs/1607.03735}{{\tt
  arXiv:1607.03735 [hep-th]}}.

\bibitem{Achucarro:2018ngj}
A.~Ach\'ucarro, S.~C\'espedes, A.-C. Davis, and G.~A. Palma, ``{Constraints on
  Holographic Multifield Inflation and Models Based on the Hamilton-Jacobi
  Formalism},'' \href{http://dx.doi.org/10.1103/PhysRevLett.122.191301}{{\em
  Phys. Rev. Lett.} {\bf 122} (2019) no.~19, 191301},
  \href{http://arxiv.org/abs/1809.05341}{{\tt arXiv:1809.05341 [hep-th]}}.

\bibitem{Arkani-Hamed:2018kmz}
N.~Arkani-Hamed, D.~Baumann, H.~Lee, and G.~L. Pimentel, ``{The Cosmological
  Bootstrap: Inflationary Correlators from Symmetries and Singularities},''
  \href{http://dx.doi.org/10.1007/JHEP04(2020)105}{{\em JHEP} {\bf 04} (2020)
  105}, \href{http://arxiv.org/abs/1811.00024}{{\tt arXiv:1811.00024
  [hep-th]}}.

\bibitem{Celoria:2021vjw}
M.~Celoria, P.~Creminelli, G.~Tambalo, and V.~Yingcharoenrat, ``{Beyond
  perturbation theory in inflation},''
  \href{http://dx.doi.org/10.1088/1475-7516/2021/06/051}{{\em JCAP} {\bf 06}
  (2021)  051}, \href{http://arxiv.org/abs/2103.09244}{{\tt arXiv:2103.09244
  [hep-th]}}.

\bibitem{Flauger:2016idt}
R.~Flauger, M.~Mirbabayi, L.~Senatore, and E.~Silverstein, ``{Productive
  Interactions: heavy particles and non-Gaussianity},''
  \href{http://dx.doi.org/10.1088/1475-7516/2017/10/058}{{\em JCAP} {\bf 10}
  (2017)  058}, \href{http://arxiv.org/abs/1606.00513}{{\tt arXiv:1606.00513
  [hep-th]}}.

\bibitem{Chen:2018brw}
X.~Chen, G.~A. Palma, B.~Scheihing~Hitschfeld, and S.~Sypsas, ``{Reconstructing
  the Inflationary Landscape with Cosmological Data},''
  \href{http://dx.doi.org/10.1103/PhysRevLett.121.161302}{{\em Phys. Rev.
  Lett.} {\bf 121} (2018) no.~16, 161302},
  \href{http://arxiv.org/abs/1806.05202}{{\tt arXiv:1806.05202 [astro-ph.CO]}}.

\bibitem{Chen:2018uul}
X.~Chen, G.~A. Palma, W.~Riquelme, B.~Scheihing~Hitschfeld, and S.~Sypsas,
  ``{Landscape tomography through primordial non-Gaussianity},''
  \href{http://dx.doi.org/10.1103/PhysRevD.98.083528}{{\em Phys. Rev. D} {\bf
  98} (2018) no.~8, 083528}, \href{http://arxiv.org/abs/1804.07315}{{\tt
  arXiv:1804.07315 [hep-th]}}.

\bibitem{Palma:2019lpt}
G.~A. Palma, B.~Scheihing~Hitschfeld, and S.~Sypsas, ``{Non-Gaussian CMB and
  LSS statistics beyond polyspectra},''
  \href{http://dx.doi.org/10.1088/1475-7516/2020/02/027}{{\em JCAP} {\bf 02}
  (2020)  027}, \href{http://arxiv.org/abs/1907.05332}{{\tt arXiv:1907.05332
  [astro-ph.CO]}}.

\bibitem{Hooshangi:2021ubn}
S.~Hooshangi, M.~H. Namjoo, and M.~Noorbala, ``{Rare Events Are
  Nonperturbative: Primordial Black Holes From Heavy-Tailed Distributions},''
  \href{http://arxiv.org/abs/2112.04520}{{\tt arXiv:2112.04520 [astro-ph.CO]}}.

\bibitem{Cai:2021zsp}
Y.-F. Cai, X.-H. Ma, M.~Sasaki, D.-G. Wang, and Z.~Zhou, ``{One Small Step for
  an Inflaton, One Giant Leap for Inflation: a novel non-Gaussian tail and
  primordial black holes},'' \href{http://arxiv.org/abs/2112.13836}{{\tt
  arXiv:2112.13836 [astro-ph.CO]}}.

\bibitem{Carr:2020xqk}
B.~Carr and F.~Kuhnel, ``{Primordial Black Holes as Dark Matter: Recent
  Developments},''
  \href{http://dx.doi.org/10.1146/annurev-nucl-050520-125911}{{\em Ann. Rev.
  Nucl. Part. Sci.} {\bf 70} (2020)  355--394},
  \href{http://arxiv.org/abs/2006.02838}{{\tt arXiv:2006.02838 [astro-ph.CO]}}.

\bibitem{Green:2020jor}
A.~M. Green and B.~J. Kavanagh, ``{Primordial Black Holes as a dark matter
  candidate},'' \href{http://dx.doi.org/10.1088/1361-6471/abc534}{{\em J. Phys.
  G} {\bf 48} (2021) no.~4, 043001},
  \href{http://arxiv.org/abs/2007.10722}{{\tt arXiv:2007.10722 [astro-ph.CO]}}.

\bibitem{Franciolini:2018vbk}
G.~Franciolini, A.~Kehagias, S.~Matarrese, and A.~Riotto, ``{Primordial Black
  Holes from Inflation and non-Gaussianity},''
  \href{http://dx.doi.org/10.1088/1475-7516/2018/03/016}{{\em JCAP} {\bf 03}
  (2018)  016}, \href{http://arxiv.org/abs/1801.09415}{{\tt arXiv:1801.09415
  [astro-ph.CO]}}.

\bibitem{Atal:2018neu}
V.~Atal and C.~Germani, ``{The role of non-gaussianities in Primordial Black
  Hole formation},'' \href{http://dx.doi.org/10.1016/j.dark.2019.100275}{{\em
  Phys. Dark Univ.} {\bf 24} (2019)  100275},
  \href{http://arxiv.org/abs/1811.07857}{{\tt arXiv:1811.07857 [astro-ph.CO]}}.

\bibitem{Musco:2020jjb}
I.~Musco, V.~De~Luca, G.~Franciolini, and A.~Riotto, ``{Threshold for
  primordial black holes. II. A simple analytic prescription},''
  \href{http://dx.doi.org/10.1103/PhysRevD.103.063538}{{\em Phys. Rev. D} {\bf
  103} (2021) no.~6, 063538}, \href{http://arxiv.org/abs/2011.03014}{{\tt
  arXiv:2011.03014 [astro-ph.CO]}}.

\bibitem{Kitajima:2021fpq}
N.~Kitajima, Y.~Tada, S.~Yokoyama, and C.-M. Yoo, ``{Primordial black holes in
  peak theory with a non-Gaussian tail},''
  \href{http://dx.doi.org/10.1088/1475-7516/2021/10/053}{{\em JCAP} {\bf 10}
  (2021)  053}, \href{http://arxiv.org/abs/2109.00791}{{\tt arXiv:2109.00791
  [astro-ph.CO]}}.

\bibitem{Starobinsky:1986fx}
A.~A. Starobinsky, ``{Stochastic de~Sitter (inflationary) stage in the early
  universe},'' \href{http://dx.doi.org/10.1007/3-540-16452-9_6}{{\em Lect.
  Notes Phys.} {\bf 246} (1986)  107--126}.

\bibitem{Goncharov:1987ir}
A.~S. Goncharov, A.~D. Linde, and V.~F. Mukhanov, ``{The Global Structure of
  the Inflationary Universe},''
  \href{http://dx.doi.org/10.1142/S0217751X87000211}{{\em Int. J. Mod. Phys. A}
  {\bf 2} (1987)  561--591}.

\bibitem{Salopek:1990re}
D.~S. Salopek and J.~R. Bond, ``{Stochastic inflation and nonlinear gravity},''
  \href{http://dx.doi.org/10.1103/PhysRevD.43.1005}{{\em Phys. Rev. D} {\bf 43}
  (1991)  1005--1031}.

\bibitem{Starobinsky:1994bd}
A.~A. Starobinsky and J.~Yokoyama, ``{Equilibrium state of a selfinteracting
  scalar field in the De Sitter background},''
  \href{http://dx.doi.org/10.1103/PhysRevD.50.6357}{{\em Phys. Rev. D} {\bf 50}
  (1994)  6357--6368}, \href{http://arxiv.org/abs/astro-ph/9407016}{{\tt
  arXiv:astro-ph/9407016}}.

\bibitem{Tsamis:2005hd}
N.~C. Tsamis and R.~P. Woodard, ``{Stochastic quantum gravitational
  inflation},'' \href{http://dx.doi.org/10.1016/j.nuclphysb.2005.06.031}{{\em
  Nucl. Phys. B} {\bf 724} (2005)  295--328},
  \href{http://arxiv.org/abs/gr-qc/0505115}{{\tt arXiv:gr-qc/0505115}}.

\bibitem{Tolley:2008na}
A.~J. Tolley and M.~Wyman, ``{Stochastic Inflation Revisited: Non-Slow Roll
  Statistics and DBI Inflation},''
  \href{http://dx.doi.org/10.1088/1475-7516/2008/04/028}{{\em JCAP} {\bf 04}
  (2008)  028}, \href{http://arxiv.org/abs/0801.1854}{{\tt arXiv:0801.1854
  [hep-th]}}.

\bibitem{Finelli:2008zg}
F.~Finelli, G.~Marozzi, A.~A. Starobinsky, G.~P. Vacca, and G.~Venturi,
  ``{Generation of fluctuations during inflation: Comparison of stochastic and
  field-theoretic approaches},''
  \href{http://dx.doi.org/10.1103/PhysRevD.79.044007}{{\em Phys. Rev. D} {\bf
  79} (2009)  044007}, \href{http://arxiv.org/abs/0808.1786}{{\tt
  arXiv:0808.1786 [hep-th]}}.

\bibitem{Finelli:2010sh}
F.~Finelli, G.~Marozzi, A.~A. Starobinsky, G.~P. Vacca, and G.~Venturi,
  ``{Stochastic growth of quantum fluctuations during slow-roll inflation},''
  \href{http://dx.doi.org/10.1103/PhysRevD.82.064020}{{\em Phys. Rev. D} {\bf
  82} (2010)  064020}, \href{http://arxiv.org/abs/1003.1327}{{\tt
  arXiv:1003.1327 [hep-th]}}.

\bibitem{Riotto:2011sf}
A.~Riotto and M.~S. Sloth, ``{The probability equation for the cosmological
  comoving curvature perturbation},''
  \href{http://dx.doi.org/10.1088/1475-7516/2011/10/003}{{\em JCAP} {\bf 10}
  (2011)  003}, \href{http://arxiv.org/abs/1103.5876}{{\tt arXiv:1103.5876
  [astro-ph.CO]}}.

\bibitem{PerreaultLevasseur:2013kfq}
L.~Perreault~Levasseur, ``{Lagrangian formulation of stochastic inflation:
  Langevin equations, one-loop corrections and a proposed recursive
  approach},'' \href{http://dx.doi.org/10.1103/PhysRevD.88.083537}{{\em Phys.
  Rev. D} {\bf 88} (2013) no.~8, 083537},
  \href{http://arxiv.org/abs/1304.6408}{{\tt arXiv:1304.6408 [hep-th]}}.

\bibitem{Burgess:2014eoa}
C.~P. Burgess, R.~Holman, G.~Tasinato, and M.~Williams, ``{EFT Beyond the
  Horizon: Stochastic Inflation and How Primordial Quantum Fluctuations Go
  Classical},'' \href{http://dx.doi.org/10.1007/JHEP03(2015)090}{{\em JHEP}
  {\bf 03} (2015)  090}, \href{http://arxiv.org/abs/1408.5002}{{\tt
  arXiv:1408.5002 [hep-th]}}.

\bibitem{Moss:2016uix}
I.~Moss and G.~Rigopoulos, ``{Effective long wavelength scalar dynamics in de
  Sitter},'' \href{http://dx.doi.org/10.1088/1475-7516/2017/05/009}{{\em JCAP}
  {\bf 05} (2017)  009}, \href{http://arxiv.org/abs/1611.07589}{{\tt
  arXiv:1611.07589 [gr-qc]}}.

\bibitem{Grain:2017dqa}
J.~Grain and V.~Vennin, ``{Stochastic inflation in phase space: Is slow roll a
  stochastic attractor?},''
  \href{http://dx.doi.org/10.1088/1475-7516/2017/05/045}{{\em JCAP} {\bf 05}
  (2017)  045}, \href{http://arxiv.org/abs/1703.00447}{{\tt arXiv:1703.00447
  [gr-qc]}}.

\bibitem{Gorbenko:2019rza}
V.~Gorbenko and L.~Senatore, ``{$\lambda \phi^4$ in dS},''
  \href{http://arxiv.org/abs/1911.00022}{{\tt arXiv:1911.00022 [hep-th]}}.

\bibitem{Mirbabayi:2019qtx}
M.~Mirbabayi, ``{Infrared dynamics of a light scalar field in de Sitter},''
  \href{http://dx.doi.org/10.1088/1475-7516/2020/12/006}{{\em JCAP} {\bf 12}
  (2020)  006}, \href{http://arxiv.org/abs/1911.00564}{{\tt arXiv:1911.00564
  [hep-th]}}.

\bibitem{Cohen:2020php}
T.~Cohen and D.~Green, ``{Soft de Sitter Effective Theory},''
  \href{http://dx.doi.org/10.1007/JHEP12(2020)041}{{\em JHEP} {\bf 12} (2020)
  041}, \href{http://arxiv.org/abs/2007.03693}{{\tt arXiv:2007.03693
  [hep-th]}}.

\bibitem{Pinol:2020cdp}
L.~Pinol, S.~Renaux-Petel, and Y.~Tada, ``{A manifestly covariant theory of
  multifield stochastic inflation in phase space: solving the discretisation
  ambiguity in stochastic inflation},''
  \href{http://dx.doi.org/10.1088/1475-7516/2021/04/048}{{\em JCAP} {\bf 04}
  (2021)  048}, \href{http://arxiv.org/abs/2008.07497}{{\tt arXiv:2008.07497
  [astro-ph.CO]}}.

\bibitem{Panagopoulos:2019ail}
G.~Panagopoulos and E.~Silverstein, ``{Primordial Black Holes from non-Gaussian
  tails},'' \href{http://arxiv.org/abs/1906.02827}{{\tt arXiv:1906.02827
  [hep-th]}}.

\bibitem{Ezquiaga:2019ftu}
J.~M. Ezquiaga, J.~Garc\'\i{}a-Bellido, and V.~Vennin, ``{The exponential tail
  of inflationary fluctuations: consequences for primordial black holes},''
  \href{http://dx.doi.org/10.1088/1475-7516/2020/03/029}{{\em JCAP} {\bf 03}
  (2020)  029}, \href{http://arxiv.org/abs/1912.05399}{{\tt arXiv:1912.05399
  [astro-ph.CO]}}.

\bibitem{Figueroa:2020jkf}
D.~G. Figueroa, S.~Raatikainen, S.~Rasanen, and E.~Tomberg, ``{Non-Gaussian
  Tail of the Curvature Perturbation in Stochastic Ultraslow-Roll Inflation:
  Implications for Primordial Black Hole Production},''
  \href{http://dx.doi.org/10.1103/PhysRevLett.127.101302}{{\em Phys. Rev.
  Lett.} {\bf 127} (2021) no.~10, 101302},
  \href{http://arxiv.org/abs/2012.06551}{{\tt arXiv:2012.06551 [astro-ph.CO]}}.

\bibitem{Pattison:2021oen}
C.~Pattison, V.~Vennin, D.~Wands, and H.~Assadullahi, ``{Ultra-slow-roll
  inflation with quantum diffusion},''
  \href{http://dx.doi.org/10.1088/1475-7516/2021/04/080}{{\em JCAP} {\bf 04}
  (2021)  080}, \href{http://arxiv.org/abs/2101.05741}{{\tt arXiv:2101.05741
  [astro-ph.CO]}}.

\bibitem{Figueroa:2021zah}
D.~G. Figueroa, S.~Raatikainen, S.~Rasanen, and E.~Tomberg, ``{Implications of
  stochastic effects for primordial black hole production in ultra-slow-roll
  inflation},'' \href{http://arxiv.org/abs/2111.07437}{{\tt arXiv:2111.07437
  [astro-ph.CO]}}.

\bibitem{Gordon:2000hv}
C.~Gordon, D.~Wands, B.~A. Bassett, and R.~Maartens, ``{Adiabatic and entropy
  perturbations from inflation},''
  \href{http://dx.doi.org/10.1103/PhysRevD.63.023506}{{\em Phys. Rev. D} {\bf
  63} (2000)  023506}, \href{http://arxiv.org/abs/astro-ph/0009131}{{\tt
  arXiv:astro-ph/0009131}}.

\bibitem{GrootNibbelink:2001qt}
S.~Groot~Nibbelink and B.~J.~W. van Tent, ``{Scalar perturbations during
  multiple field slow-roll inflation},''
  \href{http://dx.doi.org/10.1088/0264-9381/19/4/302}{{\em Class. Quant. Grav.}
  {\bf 19} (2002)  613--640}, \href{http://arxiv.org/abs/hep-ph/0107272}{{\tt
  arXiv:hep-ph/0107272}}.

\bibitem{Achucarro:2016fby}
A.~Ach\'ucarro, V.~Atal, C.~Germani, and G.~A. Palma, ``{Cumulative effects in
  inflation with ultra-light entropy modes},''
  \href{http://dx.doi.org/10.1088/1475-7516/2017/02/013}{{\em JCAP} {\bf 02}
  (2017)  013}, \href{http://arxiv.org/abs/1607.08609}{{\tt arXiv:1607.08609
  [astro-ph.CO]}}.

\bibitem{Achucarro:2019pux}
A.~Ach\'ucarro, E.~J. Copeland, O.~Iarygina, G.~A. Palma, D.-G. Wang, and
  Y.~Welling, ``{Shift-symmetric orbital inflation: Single field or
  multifield?},'' \href{http://dx.doi.org/10.1103/PhysRevD.102.021302}{{\em
  Phys. Rev. D} {\bf 102} (2020) no.~2, 021302},
  \href{http://arxiv.org/abs/1901.03657}{{\tt arXiv:1901.03657 [astro-ph.CO]}}.

\bibitem{Welling:2019bib}
Y.~Welling, ``{Simple, exact model of quasisingle field inflation},''
  \href{http://dx.doi.org/10.1103/PhysRevD.101.063535}{{\em Phys. Rev. D} {\bf
  101} (2020) no.~6, 063535}, \href{http://arxiv.org/abs/1907.02951}{{\tt
  arXiv:1907.02951 [astro-ph.CO]}}.

\bibitem{Achucarro:2019mea}
A.~Ach\'ucarro and Y.~Welling, ``{Orbital Inflation: inflating along an angular
  isometry of field space},'' \href{http://arxiv.org/abs/1907.02020}{{\tt
  arXiv:1907.02020 [hep-th]}}.

\bibitem{Woodard:2005cw}
R.~P. Woodard, ``{A Leading logarithm approximation for inflationary quantum
  field theory},''
  \href{http://dx.doi.org/10.1016/j.nuclphysbps.2005.04.056}{{\em Nucl. Phys. B
  Proc. Suppl.} {\bf 148} (2005)  108--119},
  \href{http://arxiv.org/abs/astro-ph/0502556}{{\tt arXiv:astro-ph/0502556}}.

\bibitem{van1992stochastic}
N.~G. {van Kampen}, {\em {Stochastic Processes in Physics and Chemistry}}.
\newblock 1992.

\bibitem{VanKampen:1985}
N.~G. Van~Kampen,
  \href{http://dx.doi.org/10.1016/0370-1573(85)90002-X}{``{Elimination of fast
  variables},''{\em Physics Reports} {\bf {124}} (jul, 1985)  69--160}.

\bibitem{Achucarro:2010jv}
A.~Achucarro, J.-O. Gong, S.~Hardeman, G.~A. Palma, and S.~P. Patil, ``{Mass
  hierarchies and non-decoupling in multi-scalar field dynamics},''
  \href{http://dx.doi.org/10.1103/PhysRevD.84.043502}{{\em Phys. Rev. D} {\bf
  84} (2011)  043502}, \href{http://arxiv.org/abs/1005.3848}{{\tt
  arXiv:1005.3848 [hep-th]}}.

\bibitem{Achucarro:2012sm}
A.~Achucarro, J.-O. Gong, S.~Hardeman, G.~A. Palma, and S.~P. Patil,
  ``{Effective theories of single field inflation when heavy fields matter},''
  \href{http://dx.doi.org/10.1007/JHEP05(2012)066}{{\em JHEP} {\bf 05} (2012)
  066}, \href{http://arxiv.org/abs/1201.6342}{{\tt arXiv:1201.6342 [hep-th]}}.

\bibitem{Garcia-Saenz:2019njm}
S.~Garcia-Saenz, L.~Pinol, and S.~Renaux-Petel, ``{Revisiting non-Gaussianity
  in multifield inflation with curved field space},''
  \href{http://dx.doi.org/10.1007/JHEP01(2020)073}{{\em JHEP} {\bf 01} (2020)
  073}, \href{http://arxiv.org/abs/1907.10403}{{\tt arXiv:1907.10403
  [hep-th]}}.

\bibitem{Cohen:2021fzf}
T.~Cohen, D.~Green, A.~Premkumar, and A.~Ridgway, ``{Stochastic Inflation at
  NNLO},'' \href{http://arxiv.org/abs/2106.09728}{{\tt arXiv:2106.09728
  [hep-th]}}.

\bibitem{Mirbabayi:2020vyt}
M.~Mirbabayi, ``{Markovian Dynamics in de Sitter},''
  \href{http://arxiv.org/abs/2010.06604}{{\tt arXiv:2010.06604 [hep-th]}}.

\bibitem{Witten:2010cx}
E.~Witten, ``{Analytic Continuation Of Chern-Simons Theory},'' {\em AMS/IP
  Stud. Adv. Math.} {\bf 50} (2011)  347--446,
  \href{http://arxiv.org/abs/1001.2933}{{\tt arXiv:1001.2933 [hep-th]}}.

\bibitem{Feldbrugge:2017kzv}
J.~Feldbrugge, J.-L. Lehners, and N.~Turok, ``{Lorentzian Quantum Cosmology},''
  \href{http://dx.doi.org/10.1103/PhysRevD.95.103508}{{\em Phys. Rev. D} {\bf
  95} (2017) no.~10, 103508}, \href{http://arxiv.org/abs/1703.02076}{{\tt
  arXiv:1703.02076 [hep-th]}}.

\bibitem{Serone:2017nmd}
M.~Serone, G.~Spada, and G.~Villadoro, ``{The Power of Perturbation Theory},''
  \href{http://dx.doi.org/10.1007/JHEP05(2017)056}{{\em JHEP} {\bf 05} (2017)
  056}, \href{http://arxiv.org/abs/1702.04148}{{\tt arXiv:1702.04148
  [hep-th]}}.

\bibitem{Cohen:2021jbo}
T.~Cohen, D.~Green, and A.~Premkumar, ``{A Tail of Eternal Inflation},''
  \href{http://arxiv.org/abs/2111.09332}{{\tt arXiv:2111.09332 [hep-th]}}.

\bibitem{Baumgart:2019clc}
M.~Baumgart and R.~Sundrum, ``{De Sitter Diagrammar and the Resummation of
  Time},'' \href{http://dx.doi.org/10.1007/JHEP07(2020)119}{{\em JHEP} {\bf 07}
  (2020)  119}, \href{http://arxiv.org/abs/1912.09502}{{\tt arXiv:1912.09502
  [hep-th]}}.

\bibitem{Assadullahi:2016gkk}
H.~Assadullahi, H.~Firouzjahi, M.~Noorbala, V.~Vennin, and D.~Wands,
  ``{Multiple Fields in Stochastic Inflation},''
  \href{http://dx.doi.org/10.1088/1475-7516/2016/06/043}{{\em JCAP} {\bf 06}
  (2016)  043}, \href{http://arxiv.org/abs/1604.04502}{{\tt arXiv:1604.04502
  [hep-th]}}.

\bibitem{Vennin:2016wnk}
V.~Vennin, H.~Assadullahi, H.~Firouzjahi, M.~Noorbala, and D.~Wands,
  ``{Critical Number of Fields in Stochastic Inflation},''
  \href{http://dx.doi.org/10.1103/PhysRevLett.118.031301}{{\em Phys. Rev.
  Lett.} {\bf 118} (2017) no.~3, 031301},
  \href{http://arxiv.org/abs/1604.06017}{{\tt arXiv:1604.06017 [astro-ph.CO]}}.

\bibitem{Salopek:1990jq}
D.~S. Salopek and J.~R. Bond, ``{Nonlinear evolution of long wavelength metric
  fluctuations in inflationary models},''
  \href{http://dx.doi.org/10.1103/PhysRevD.42.3936}{{\em Phys. Rev. D} {\bf 42}
  (1990)  3936--3962}.

\bibitem{Palma:2020ejf}
G.~A. Palma, S.~Sypsas, and C.~Zenteno, ``{Seeding primordial black holes in
  multifield inflation},''
  \href{http://dx.doi.org/10.1103/PhysRevLett.125.121301}{{\em Phys. Rev.
  Lett.} {\bf 125} (2020) no.~12, 121301},
  \href{http://arxiv.org/abs/2004.06106}{{\tt arXiv:2004.06106 [astro-ph.CO]}}.

\bibitem{Fumagalli:2020adf}
J.~Fumagalli, S.~Renaux-Petel, J.~W. Ronayne, and L.~T. Witkowski, ``{Turning
  in the landscape: a new mechanism for generating Primordial Black Holes},''
  \href{http://arxiv.org/abs/2004.08369}{{\tt arXiv:2004.08369 [hep-th]}}.

\bibitem{Fumagalli:2020nvq}
J.~Fumagalli, S.~Renaux-Petel, and L.~T. Witkowski, ``{Oscillations in the
  stochastic gravitational wave background from sharp features and particle
  production during inflation},''
  \href{http://dx.doi.org/10.1088/1475-7516/2021/08/030}{{\em JCAP} {\bf 08}
  (2021)  030}, \href{http://arxiv.org/abs/2012.02761}{{\tt arXiv:2012.02761
  [astro-ph.CO]}}.

\bibitem{Ballesteros:2021fsp}
G.~Ballesteros, S.~C\'espedes, and L.~Santoni, ``{Large power spectrum and
  primordial black holes in the effective theory of inflation},''
  \href{http://arxiv.org/abs/2109.00567}{{\tt arXiv:2109.00567 [hep-th]}}.

\bibitem{Panagopoulos:2020sxp}
G.~Panagopoulos and E.~Silverstein, ``{Multipoint correlators in multifield
  cosmology},'' \href{http://arxiv.org/abs/2003.05883}{{\tt arXiv:2003.05883
  [hep-th]}}.

\bibitem{Winitzki:1999ve}
S.~Winitzki and A.~Vilenkin, ``{Effective noise in stochastic description of
  inflation},'' \href{http://dx.doi.org/10.1103/PhysRevD.61.084008}{{\em Phys.
  Rev. D} {\bf 61} (2000)  084008},
  \href{http://arxiv.org/abs/gr-qc/9911029}{{\tt arXiv:gr-qc/9911029}}.

\bibitem{Creminelli:2008es}
P.~Creminelli, S.~Dubovsky, A.~Nicolis, L.~Senatore, and M.~Zaldarriaga, ``{The
  Phase Transition to Slow-roll Eternal Inflation},''
  \href{http://dx.doi.org/10.1088/1126-6708/2008/09/036}{{\em JHEP} {\bf 09}
  (2008)  036}, \href{http://arxiv.org/abs/0802.1067}{{\tt arXiv:0802.1067
  [hep-th]}}.

\bibitem{Cespedes:2020xqq}
S.~C\'espedes, A.-C. Davis, and S.~Melville, ``{On the time evolution of
  cosmological correlators},''
  \href{http://dx.doi.org/10.1007/JHEP02(2021)012}{{\em JHEP} {\bf 02} (2021)
  012}, \href{http://arxiv.org/abs/2009.07874}{{\tt arXiv:2009.07874
  [hep-th]}}.

\end{thebibliography}\endgroup

\end{document}